\documentclass[fleqn,usenatbib]{mnras}
\usepackage{graphicx}
\usepackage{aas_macros}
\usepackage{amssymb}
\usepackage{url}

\title[NuSTAR and Parkes observations of the transitional millisecond pulsar 
binary  XSSJ12270-4859]
{NuSTAR and Parkes observations of the transitional millisecond pulsar binary  
XSS\,J12270-4859 in the rotation-powered state
}
\author[D. de Martino et al.]{D.~de Martino$^{1}$, \thanks{E-mail: domitilla.demartino@inaf.it}
A. Papitto$^{2}$, 
M.~Burgay$^{3}$, 
A.~Possenti$^{3}$,
F.~Coti Zelati$^{4}$,
\newauthor
N.~Rea$^{4}$,
D.F.~Torres$^{4,5,6}$,
T.M.~Belloni$^{7}$\\ 
$^{1}$ INAF $-$ Osservatorio Astronomico di Capodimonte, Salita Moiariello 16, I-80131 Napoli, Italy\\
$^{2}$ INAF $-$ Osservatorio Astronomico di Roma, via Frascati 33, I-00076, Monteporzio Catone (RM), Italy\\
$^{3}$ INAF $-$ Osservatorio Astronomico di Cagliari, Via della Scienza, I-09047
Serlagius (CA), Italy\\
$^{4}$ Instituci\'o Catalana de Recerca i Estudis Avan\c{c}ats (ICREA), E-08010, Barcelona, Spain\\
$^{5}$ Institute of Space Sciences (ICE, CSIC), Campus UAB, Carrer de Can Magrans, E-08193, Barcelona, Spain\\
$^{6}$ Institut d'Estudis Espacials de Catalunya (IEEC), E-08034 Barcelona, Spain\\
$^{7}$  INAF $-$  Osservatorio Astronomico di Brera, Via E. Bianchi 46, I-23807
Merate (LC), Italy
}

\begin{document}

\date{Accepted 2020 January 15. Received 2019 December 16; in original form 2019 October 31}

\pagerange{\pageref{firstpage}--\pageref{lastpage}} \pubyear{2019}

\maketitle

\label{firstpage}

\begin{abstract}

We report on the first {\em NuSTAR} observation of the transitional 
millisecond pulsar binary XSS\,J12270-4859
during its current rotation-powered state,
complemented with a 2.5\,yr-long radio monitoring at {\em
Parkes} telescope and archival {\em XMM-Newton} and {\em Swift} X-ray 
and optical data. The radio pulsar is mainly detected at 1.4\,GHz displaying
eclipses over $\sim 40\%$ of the 6.91\,h orbital cycle. We derive a
new updated radio ephemeris 
to study the 3-79\,keV light curve that displays a significant 
orbital modulation with fractional amplitude of 28$\pm 3\%$,  a structured
maximum centred at the inferior conjunction of the pulsar and no
cycle-to-cycle or low-high-flaring mode variabilities. 
The average X-ray spectrum, extending up to $\sim$70\,keV without a 
spectral break, is well described by a simple power-law with photon 
index $\Gamma = 1.17\pm0.08$ giving a 3-79\,keV luminosity of  
$\rm 7.6_{-0.8}^{+3.8} \times
10^{32}\,erg\,s^{-1}$, for a distance of 1.37$_{-0.15}^{+0.69}$\,kpc.
Energy resolved orbital light curves reveal
that the modulation is not energy dependent from 3\,keV to 25\,keV and
is undetected with an upper limit of $\sim$10$\%$ above 25\,keV.
Comparison with previous X-ray {\it XMM-Newton} observations 
in common energy ranges confirms that the modulation amplitudes 
vary on timescales of 
a few months, indicative of a non-stationary contribution 
of the intrabinary shock formed by the colliding winds of the pulsar 
and the companion. A more detailed inspection
of energy resolved modulations than previously reported gives hints
of a mild softening at superior conjunction of the pulsar below 3\,keV,
likely due to the contribution of the thermal emission 
from the neutron star. 
The intrabinary shock emission, if extending into the MeV range, 
would be energetically capable alone to irradiate the donor star. 

\end{abstract}


\section{Introduction}

Millisecond pulsars (MSPs) are old neutron stars (NSs) in close binary systems
which were spun-up to very short periods during a previous
Gyr-long phase of mass accretion from an evolved companion. According to the recycling
scenario \citep{Alpar82,Backer82} during the accretion phase, MSP binaries were 
Low-Mass X-ray Binaries (LMXBs) and turned into radio and Gamma-ray pulsars when mass
accretion ceased.  The first evidence of a transition between the
two states was discovered in the radio MSP binary PSR\,J1023+0038 that was  
found in an accretion disc-state between 2001-2004 
\citep{Archibald09} and entered again in a LMXB state in 2013 
\citep{Stappers14,Patruno14,Bogdanov15}. 
Similar transitions were also observed in IGR\,J1825-2452, in the 
M28 globular cluster \citep{Papitto13} and in the galactic
field X-ray source XSS\,J12270-4859 (henceforth J1227) 
\citep[][]{Bassa14,Bogdanov14,deMartino14}. The transitions  
occured on timescales much shorter than secular evolution, likely controlled 
by variations in the mass transfer rate from the late-type companion. 
These three systems, dubbed transitional MSPs (tMSPs), 
harbour non-degenerate companions in tight (5-12\,h) orbits and fall in 
the increasing class of redbacks (RB) (14 confirmed
so far with $\rm M_2 \sim 0.2-0.4\,M_{\odot}$) as opposed
to those dubbed black widows (BW) containing degenerate companions 
with $\rm M_2 << 0.1\,M_{\odot}$ \citep{Roberts15,Strader19}. 

Transitional systems are enigmatic and complex binaries. During the LMXB state they display peculiar behaviour from 
radio, optical up to X-ray and Gamma-rays \citep{deMartino10,deMartino13,Archibald15,
Papitto15,Ambrosino17,Bogdanov18,Papitto19}. Only IGR\,J1825-2452  displayed an 
outburst \citep{Papitto13}, while
the other two were never recorded in such state. When in the disc-state, 
tMSPs  are characterised by a 
subluminous X-ray emission ($\rm L_X\sim 10^{33} - 10^{34}\,\rm erg\,s^{-1}$)  
with  high, low and flare  ``modes''  \citep{linares14}. In IGR\,J1825-2452 and 
PSR\,J1023+0038, the X-ray low-modes  were found to be accompained by radio 
flares in quasi-simultaneous observations 
\citep{Ferrigno14,Bogdanov18}, possibly due to outflowing material. 
Additionally, the  
presence of X-ray pulses during high X-ray modes in PSRJ1023+0038 \citep{Archibald15}
and J1227 \citep{Papitto15} was interpreted as signature of accretion onto the NS. 
This interpretation is
challenged by the detection of optical pulsations \citep{Ambrosino17} that cannot
be ascribed to accretion. The optical and X-ray 
pulsations were recently detected simultaneously during the X-ray high
modes \citep{Papitto19}. Instead, during flares lower amplitude optical 
pulses were also detected but not in the X-rays.

\noindent A variety of interpretation have been proposed for the complex behaviour during the 
LMXB state: an  enshrouded pulsar \citep{CotiZelati14,Takata14},
a pulsar in a propeller state \citep{Papitto14,Papitto_Torres15}, 
an intermittent propellering radio pulsar \citep{Ertan17} 
or a low accretion rate pulsar from a trapped disc near corotation
\citep{Dangelo12} (see also \cite{Campana_Disalvo18}, for a review).  
Whether the disc shocks with a striped pulsar wind  at a 
few light cylinder radii away from the pulsar, giving rise synchrotron emission 
producing the optical and X-ray pulses, 
is a challenging intepretion put forward by \cite{Papitto19} 
\citep[see also][]{Veledina19,Campana19}.

In the rotation-powered (RMSP) state transitional systems 
behave as all other RB binaries \citep{Roberts15,Roberts18}, displaying
long radio eclipses, up to 60$\%$ of the orbit, due to the passage of the NS through ionised material surrounding 
the companion, likely in the form of an intrabinary  shock (IBS) produced by 
the interaction of the pulsar
wind with that of the late-type star. The shock is also expected to emit a non-thermal spectrum and to produce a large
X-ray orbital modulation \citep{Arons_Tavani93}. Such modulations are 
indeed observed in both RBs  and BWs,
with the difference that the X-ray orbital maximum occurs at the inferior conjunction
of the NS in the former group and 
viceversa in the latter. To explain the opposite phasing of the X-ray maxima
and since RBs are expected to possess stronger winds than the BWs, 
the shock would be oriented towards the pulsar in the RBs but towards the
companion star in BWs \citep[]{Roberts15,Romani16,Wadiasingh17}.

When in RMSP state, X-ray msec pulses were observed in PSRJ1023+0038  
at a few percent ($\sim 10\%$, rms) level \citep{Archibald10} and 
remain still undetected in J1227 \citep{Papitto15}.
Instead Gamma-ray pulsations were detected with {\it Fermi}-LAT in J1227 
at $5\sigma$ level nearly in phase with the high-frequency (1.4\,GHz) radio 
pulses, indicating  an origin in the pulsar magnetosphere, 
\citep{Venter12,Johnson14,Johnson15}. In PSR\,J1023+0038 only a weak evidence
of Gamma-ray pulses was found at 3.7$\sigma$ level \citep{Archibald13}.
We  also note here  that the Gamma-ray and X-ray fluxes changed  by a few, 
3-10\,times, between the LMXB and RMSP states in J1227 and 
PSRJ1023+0038, respectively \citep{Torres17}.

How transitions occur and whether all RBs could be tMSPs is a key issue to be investigated yet.  
The study of the long-term
behaviour, especially in the X-ray band where the  IBS  is expected to 
dominate, may be a powerful means to infer changes in the
 shock geometry and in turn the mass accretion rate. Also, the level of irradiation 
of the donor star is a key ingredient
in understanding whether RBs could be prone to perform transitions. In fact, 
orbital modulations in the 
optical range \citep[e.g][]{Thorstensen05,Romani15,Hui15,deMartino15,Bellm16,AlNoori18} 
reveal in some systems, such as PSR\,J1023+0038,
J1227, PSR\,J2215+51 and PSR\,J2339-0533, strong heating of the companion, 
while others, despite displaying significant 
X-ray orbital modulations, such as PSR\,J2129-0429, have companions suffering
little irradiation. 

Of the two known tMSPs in the field, PSR\,J1023+0038 is  currently in a LMXB 
state \cite[see][for latest results]{Papitto19}, while J1227 is in a radio-pulsar 
state \citep{Bassa14,Roy15}. Both systems have NS spinning at similar
periods (1.69\,ms). After the transition to the RMSP state occurred in 
late 2012-early 2013, the soft (0.3-10\,keV) X-ray emission of J1227 was
studied by \cite{Bogdanov14} and \cite{deMartino15}.
The spectrum was  found to be non-thermal with a power law photon index 
$\Gamma \sim 1.1$, harder
than that ($\sim$1.7) observed during the LMXB state 
\citep{deMartino10,deMartino13}. 
The X-ray emission was variable at the 6.91\,h orbital period. 
Comparison of the two observations peformed in Dec.\,2013 and June\,2014, 
showed that the amplitude of the orbital modulation changed by a factor of $\sim$2 
\citep{deMartino15}.  
This variability was surprisingly anti-correlated with the 
simultaneous optical U-band light
curve,  where instead the modulation decreased by a factor of two.
This suggested a variability in the intrabinary shock over a timescale
of a few months.
The hard spectral shape in the soft range is similar to other RBs, with 
a few also  observed above 10\,keV and  detected up to 40-70\,keV
without displaying a spectral break
\citep[][]{tendulkar14,kong17,AlNoori18,Kandel19}. 
   
We here present the first hard X-ray study of J1227 during its current 
rotation-powered state based
on an observation performed with the {\it NuSTAR} satellite. We complement
the analysis using previous {\it XMM-Newton} and archival Neil Gehrels 
{\em Swift} (henceforth {\em Swift}) 
data for a comparison among different epochs. We also report on the
radio mo\-ni\-to\-ring programme conducted at the {\it Parkes} telescope over 
2.5\,yrs that also gives a contemporaneous coverage with the {\it NuSTAR} 
observation.
In Sect.\,2 we report the radio observations and analysis. In Sect.\,3 
the X-ray observation are described  and the timing and spectrscopic analyses
are reported in Sect.\,4 and 5, respectively. The results are discussed in
Sect.\,6.

\section{The radio monitoring}

J1227 was observed monthly at the {\it Parkes} 64-m radio
telescope from June 2014 to February 2017, as part of the project
P880. Data were acquired mainly around 1.4 GHz with the central beam
of the 20-cm multibeam receiver \citep{swb+96} or, when unavailable,
with the H-OH receiver\footnote{ 
\url{https://www.parkes.atnf.csiro.au/observing/documentation/}
}. 
A small number of data were obtained simultaneously around 3.1\,GHz (10-cm) 
and 0.7\,GHz (50-cm) using the coaxial 1050\,cm receiver. 
Most observations were carried
out in search mode (with folding of the data being done off-line to
correct for possible orbital period variations) using either the BPSR
digital signal processor \citep{kjv+10} or the
DFB4\footnote{ \url{http://www.jb.man.ac.uk/pulsar/observing/DFB.pdf}} 
and, in parallel, CASPSR, the CASPER {\it Parkes} Swinburne
Recorder\footnote{\url{http://astronomy.swin.edu.au/pulsar/?topic=caspsr}}
in folding mode and applying coherently dedispersion. Table
\ref{tab:radio} summarises the main  parameters.

\begin{table}
\begin{center}
\caption{Main parameters of the observing systems adopted at the
  Parkes telescope. The backend name is listed in column 1, while
  columns 2 to 5 report the central frequency, the bandwidth and the
  number of frequency channels used. For search mode observations, column 5
  reports the sampling time in $\mu$s, while, for folding mode
  ones, the number of profile bins.}
\label{tab:radio}
\begin{tabular}{lcccl}
\hline
Backend & $\nu_c$ & Bandwith & N$_{\rm{chan}}$ & $\rm t_{samp}$/N$_{\rm bin}$ \\
        & (MHz)   & (MHz)    & & \\
\hline
BPSR   & 1352 &  340 &  870     & 64 $\mu$s \\ 
DFB4   & 1369 &  256 & 1024     & 100 $\mu$s \\
       & 1369 &  256 & 512/1024 & 256/512 \\
       & 3100 & 1024 & 512      & 128 $\mu$s \\
       & 3100 & 1024 & 1024     & 256 \\
CASPSR & 1352 &  340 & 435      & 128 \\ 
       &  728 &   64 &  82      & 128 \\
\hline
\end{tabular}
\end{center}
\end{table}

\begin{figure}
\begin{center}
{\includegraphics[width=9cm,angle=0]{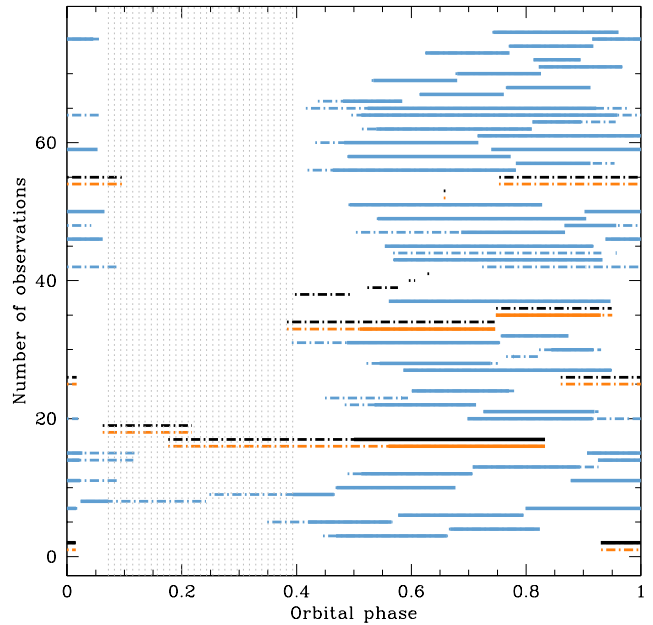}}
\caption{Summary of all the observation
    (dashed lines) and detections (solid lines) performed at Parkes as
    part of P880. The observations at 20\,cm  (1.4\,GHz) are plotted in
     light blue, those at 10\,cm (3.1\,GHz) in  black  and those 
at 50\,cm  (0.7\,GHz) in  orange. The grey
    dotted area marks the orbital phases where the pulsar was never
    detected.}
\label{fig:radio_ecli}
\end{center}

\end{figure}

The observations were planned in such a way as to avoid those orbital
phases in which the pulsar was likely to be completely eclipsed, but
also to start or finish close eclipse ingress or egress, to monitor
the possible variability of the eclipse extent. The orbital phases
were predicted on the basis of orbital ephemeris constantly updated
during the observing project. Figure\,\ref{fig:radio_ecli} shows the
observations (dashed lines) and detections (solid lines) at the three
observed frequencies. 
 The test pulsar was clearly detected, excluding a malfunctioning of 
the acquisition system. 
The eclipse extends at least between phases 0.07 and 0.39, but
shows a large variability in time, with the pulsar being sometimes
undetectable for the entire orbit. The pulsar was seen only seldom at
frequencies different from 1.4 GHz:  at 0.7 GHz, radio frequency
interferences and absorption of the signal from the surrounding medium
likely play a major role in the non detections; at 3.1\,GHz it is barely
detectable possibly due to the intrinsic spectrum of the source. 
A measurement of the spectral index of J1227 with our data set is, however, 
not possible since most of the data are uncalibrated and because the source 
is highly variabile also far from  eclipses, with a signal-to-noise ratio
varying by a factor up to 5 at 1.4 GHz. 
Because of this, the measurement of the flux density is very
uncertain, preventing us to give meaningful constraints on the spectral index.
We also here recall that previous
observations at frequencies above 1.4\,GHz only provided upper limits
to the flux \citep{Bassa14}.

\subsection{The radio ephemeris}

Timing analysis was carried out on the 20-cm data using only the data 
with orbital phases outside the range between 0.05 and 0.4, in order to exclude points affected by 
extra dispersive delays due to the eclipsing material. A profile template was 
created by adding in phase several observations together (for a total of about
40 hours). One or more times of arrival (ToAs) were then created by 
cross-correlation of the standard profile with each observing epoch.
Usually a ToA was computed for every 15-minutes of observation, but in some
cases, depending on the intensity of the signal, we used data segments from
5 to 60 minutes in length. The above steps were carried out using the software
{\sc psrchive}\footnote{\url{http://psrchive.sourceforge.net/}}. The timing
analysis was done with {\sc tempo2 } \citep{hem06}. 
Table\,\ref{tab:ephemTOT} reports the radio ephemeris obtained with the
from the full 20-cm {\it Parkes}
data set spanning 2.5\,yrs. The inclusion of two orbital frequency 
derivatives was
necessary to flatten the timing residuals and thus we adopted the $BTX$ 
binary model\footnote{Binary models  are described at
\url{http://tempo.sourceforge.net/ref_man_sections/binary.txt} }
%
The planetary ephemeris DE405 \citep{sta98c} and the TT(TAI) time standard 
(e.g. \citealt{lk05}) were used.
This ephemeris is in accord within errors with those reported by \cite{Roy15}
who instead adopted the $ELL1$ binary model over a timespan of 270\,days. 
 The 
$BT$ model, including just the first derivative of the orbital period, 
was used on data-spans up to 300 days. After that, orbital trends clearly 
affected the timing residuals. The $BT$ model, however, resulted in an rms 
comparable to that reported in Table 2, for data spans as long as 600 days, 
after which higher order derivatives were needed to phase connect further 
data points.

\begin{table}
{\scriptsize
\begin{center}
\caption{Timing parameters of PSR\,J1227$-$4853 obtained from the 20-cm 
{\it Parkes} data over the entire 2.5 year dataspan. Timing results are
in  Barycentric Dynamical Time (TDB)}
\label{tab:ephemTOT}
\begin{tabular}{ll}

\hline
Parameter &  Value$^a$ \\
\hline
R.A. (J2000) (h:m:s) & 12:27:58.7186(4) \\
DEC. (J2000) (d:m:s) & -48:53:42.707(5) \\
Pulsar frequency $f$ (Hz) & 592.98777342489(14) \\
Frequency derivative $\dot f$ (Hz\,s$^{-1}$) & -4.619(11)$\times10^{-15}$ \\
Period epoch (MJD) & 57180 \\
Dispersion measure (pc\,cm$^{-3}$) & 43.423(3) \\
Binary model & BTX \\
Orbital period $\rm P_b$ (d) & 0.287887802(5) \\
Orbital frequency $f_b$ (Hz) & 4.02034315(11)$\times10^{-5}$ \\
Orbital frequency derivative $\rm \dot f_b$(s$^{-2}$)  & 3.74(6)$\times10^{-18}$ \\
Orbital frequency second derivative $\rm \ddot f_b$(s$^{-3}$)  & -7.77(11)$\times10^{-26}$ \\
Epoch of NS ascending node T$_{\rm asc}$ (MJD) & 56700.907021(4) \\
Projected semimajor axis $\rm a1$ (lt-s)  & 0.668492(14) \\
Span of timing data (MJD) & 56824.257-57685.210 \\
Nuber of TOAs & 593 \\
Post-fit residuals rms ($\mu$s) & 49.858 \\
\hline
\end{tabular}
\end{center}
}
\begin{flushleft}
$^a$ $2\,\sigma$ errors on the last quoted digit(s)\\
\end{flushleft}
\end{table}

In order to properly fold the X-ray data presented here, we
also created a more local timing solution using approximately 170\,days
of {\it Parkes} data around the {\it NuSTAR} observation. 
Only one orbital period
derivative was needed to properly phase connect the data on this
shorter time-span and therefore we used binary model BT. 
The related ephemeris are shown in Table
\ref{tab:ephemNuSTAR}.

\begin{table}
{\scriptsize
\begin{center}
\caption{Timing ephemeris obtained from the 20-cm {\it Parkes} 
data over $\sim 170$ days around the NuSTAR observation. 
The parameters reported are as in Table
  \ref{tab:ephemTOT}} 
\label{tab:ephemNuSTAR}
\begin{tabular}{ll}
\hline
Parameter &  Value \\
\hline
R.A. (J2000) (h:m:s) & 12:27:58.7194(14) \\
DEC. (J2000) (d:m:s) & -48:53:42.712(19) \\
Pulsar frequency $f$ (Hz) & 592.9877734395(9) \\
Frequency derivative $\dot f$ (Hz\,s$^{-1}$) & -4.3(3)$\times10^{-15}$ \\
Period epoch (MJD)& 57139 \\
Dispersion measure (pc\,cm$^{-3}$) & 43.423(3) \\
Binary model & BT \\
Orbital period $\rm P_b$ (d) & 0.287887065(5) \\ 
Orbital period derivative $\rm \dot P_b$  & -2.5(14)$\times10^{-10}$ \\
Epoch of NS ascending node T$_{\rm asc}$ (MJD) & 57139.0715595(6)\\
Projected semimajor axis $a1$ (lt-s)  & 0.668482(16)\\ 
Span of timing data (MJD) & 57063.530-57233.225 \\
Nuber of TOAs & 189 \\
Post-fit residuals rms ($\mu$s) & 23.097 \\
\hline
\end{tabular}
\end{center}} 
\end{table}

\subsection{Distance estimates}

The dispersion measure (DM) has been evaluated using a few measures at 50-cm
avoiding eclipses resulting in DM= 43.423(3)\,$\rm pc\,cm^{-3}$, where here
uncertainty is at 2$\sigma$ level. It has then been kept fixed in the analysis 
of the other data sets at 20-cm.  This determination, although at lower accuracy,
is in agreement with that previously derived (43.4235(7)) \citep{Roy15}.
The DM value, when adopting \cite{cordes02} model of the Galactic 
electron 
density distribution, gives a distance of 1.4\,kpc \citep{Roy15}. However
as noted by \cite{jennings18} the DM-based distances calculated using 
either \cite{cordes02} or \cite{yao17} models are on average slightly 
underestimated.  For J1227 there is no parallax measurement from radio 
observations yet. However,   
the recent release of {\it Gaia } DR2 parallaxes 
\citep{gaia18a} allows to obtain first direct distance measures for MSP binaries
with relatively bright optical companions.
J1227 has a  parallax measure $\tilde \pi$=0.624$\pm$0.168\,mas which, 
accounting
for the DR2 parallax average zero-point offset of -0.029\,mas 
\citep[see][]{lindegren18}, translates into 
a distance $\rm d_{\tilde \pi}$=1.53$\pm$0.39\,kpc. However reliable distance 
estimates
should account for the space density distribution of the objects. Adopting 
a weak distance prior that varies as a function of Galactc longitude and latitude
according to the Galactic model described in \citep{bailer-Jones18}\footnote
{\url{http://gaia.ari.uni-heidelberg.de/tap.html}}, we estimate a distance of 
$\rm D_{BJ}=1.51_{-0.35}^{+0.59}$\,kpc, taking into account the average 
zero-point
offset. On the other hand adopting a distance prior based on the 
Galactic pulsar population of \cite{lorimer06} and also accounting for 
the average zero-point offset,  \cite{jennings18} derive  
$\rm D_L$=1.37$_{-0.15}^{+0.69}$\,kpc. The two estimates agree within their
1$\sigma$ uncertainties. We then adopt the latter distance 
for J1227. \\

 We also estimate the intrinsic spin-down power, 
$\rm \dot E=4\,\pi^2\,I\,\dot P/P^3$, where P and $\rm \dot P$ are the spin period
 and its derivative  and 
$\rm I \simeq (M_{NS}/1.4\,M_{\odot})\, (R_{NS}/10\,km)\, 10^{45}\,g\,cm^{-2}$, 
is the momentum of inertia of the NS.  We use the refined pulsar spin frequency
and first derivative (Table\,\ref{tab:ephemTOT}) that are more accurate than those
derived by \cite{Roy15}, which were based on a shorter, 270\,d,  timespan. 
With the accurate  {\it Gaia} DR2 proper motion ($\rm \mu = 
20.13(23)\,mas\,yr^{-1}$),
we correct  the observed $\rm \dot P$ for the Shklovskii effect: 
$\rm \dot P_{Sh}/P = v_t^2/c\,D$, where $\rm v_t$ is the transverse velocity.
For a distance of 1.37\,kpc we derive 
$\rm \dot P_{corr} = \dot P_{obs} - \dot P_{Sh} = 1.086(2)\times10^{-20}$.  
Correspondingly, adopting
a NS with 1.4\,$\rm M_{\odot}$ and 10\,km radius, we derive
$\rm \dot E=8.94(2)\times10^{34}\,erg\,s^{-1}$.  As a comparison, \cite{Roy15} 
derived  $\rm \dot E= 9.0(8)\times10^{34}\,erg\,s^{-1}$.  However, 
allowing the full range of uncertainties
in the distance $\rm D_L$, we obtain 
$\rm \dot E=8.9^{+0.2}_{-0.9}\times10^{34}\,erg\,s^{-1}$.

\section{The NuSTAR Observation}

J1227 was observed by {\em NuSTAR} \citep{Harrison13}
 from Apr. 26 to 28, 2015 with both  telescope modules 
FPMA and FPMB, covering 171.8\,ks, for a total effective exposure time 
of 96.4\,ks (OBSID: 30101033002).  

The processing and filtering of the NuSTAR photon event
data was performed with the standard NuSTAR pipeline (NUSTARDAS) 
version v1.8.0 with calibration data CALDB 20190503. The source 
photon events were extracted from circular regions of 25\,pixels (61$^{''}$) 
radius and the
background from a region of 50\,pixels in a source-free region. 
Photon event arrival times from both FPMA and FPMB were corrected to the 
Solar System Baricentre using the JPL DE405 ephemeris using the 
radio position  reported in Table\,2. 
Combined net light curves were constructed by subtracting the source 
light curves for the 
corresponding background in each  FPMA and FPMB module and averaged together.
The source was detected at an average net count rate of 
0.037$\rm \pm0.004\,cts\,s^{-1}$ in the  3-79\,keV  range. 
Spectra from the two telescope modules
were extracted using the {\sc nuproducts} and 
then binned to achieve a minimum of
35 photons per bin. All photons below 3\,keV (channel 35) and above 79\,keV
(channel 1935) were flagged as bad.

\begin{figure*}
\includegraphics[width=2.0in, height=5.5in,angle=-90]{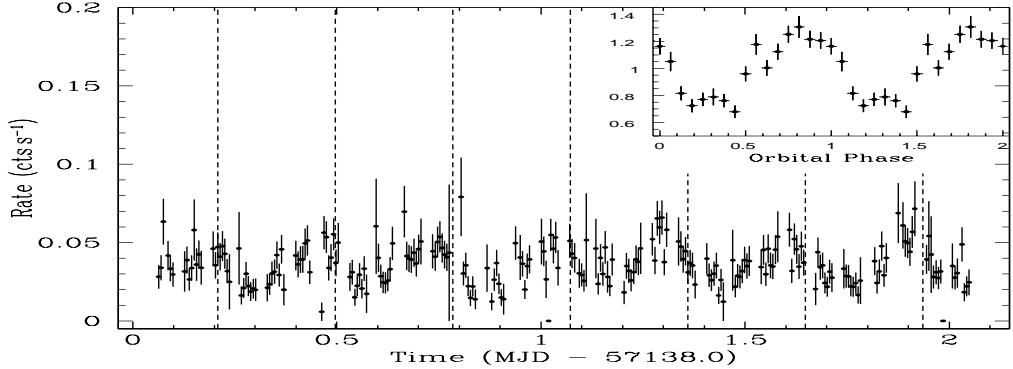}
\caption{The background subtracted {\em NuSTAR} light curve of J1227 binned at 500\,s in the 
3-79\,keV energy range.  The periodic gaps in the data are due to the 
Earth occultation of the source. The vertical dashed lines show the times 
of the passage of the pulsar at the ascending node of the orbit 
($\rm T_{asc}$) adopting the orbital radio 
ephemeris reported in Table\,\ref{tab:ephemNuSTAR}.
The inset shows the 3-79\,keV normalised light
curve folded at the 6.91\,h orbital period binned in 16 phase intervals.
Phase zero corresponds to the passage of the NS at the 
ascending node of the orbit.}
\label{nustar_lc}
\end{figure*}

\section{Timing analysis}

\subsection{The X-ray variability}

 The combined (FPMA and FPMB) net X-ray light curve in the 3-79\,keV range 
(Fig.\,\ref{nustar_lc}) 
shows variations in the count rate  over the 47\,h-long observation on 
timescale of $\sim$7\,h, consistent with the 6.91\,h orbital period.
Neither flaring activity nor sudden 
drops in the count rate are observed throughout the observation. 
A search for orbital modulation was carried out by performing Fourier analysis 
on the barycentred corrected light curve in the 3-79\,keV range adopting a  
time bin of 340\,s. 
A strong peak at $\sim$25000\,s as well as minor peaks at the beat 
frequency of the {\em NuSTAR} spacecraft are detected.
A sinusoidal fit to the light curve gives a period of 25060$\pm$235s 
(1$\sigma$ confidence level) ($\chi^2_{\nu}$=1.08, d.o.f.=319).
An epoch-folding method \citep{Leahy87} 
was also applied, adopting  8 orbital phase bins using various trial 
periods around the radio nominal period of 24873\,s with a period 
resolution of 50s. The $\chi^2$ distribution shows a peak at 9$\sigma$ 
level at a period of 25232$\pm$ 90\,s as evaluated from a Gaussian fit. 
These determinations of the period are however consistent 
within their 1$\sigma$ and 3$\sigma$ uncertainties 
with the accurate radio orbital period reported
in Tables\,2 and 3, respectively. No cycle to cycle variation is observed
in the average count rate level over the $\sim$6.9 orbital cycles covered by 
the {\em NuSTAR} observation (see Fig.\,\ref{nustar_lc}). 

The FPMA and FPMB background subtracted
light curves in different energy ranges were then merged and folded  at the 
binary orbital period $\rm P_b$=24873.4424\,s, adopting 
as phase $\rm \Phi_{orb}$=0.0 the time of passage of the pulsar
at the ascending node $\rm T_{asc}$= 57139.0715595 (MJD), 
reported in Table\,3. 
The folded light curve in the 3-79\,keV band, evalutated over
16 phase bins,   (see inset in Fig.\,\ref{nustar_lc})
shows a structured broad maximum at orbital phase $\sim$0.8 and a minimum
at $\sim$0.25, close to the inferior and superior conjunction of the
pulsar, respectively. The fractional amplitude, defined
as $\rm (F_{max} - F_{min})/(F_{max} + F_{min}$),
where $\rm F_{max}$ and $\rm F_{min}$ are the maximum and minimum count 
rates detected in the light curve, respectively,  is $28(3)\%$. 
The rise to the maximum is slower than the subsequent
decay, reaching a peak at $\rm \Phi_{orb}$=0.9. Also the
minimum is not smooth or flat-bottom but rather structured towards the rise to
the maximum. 
A well defined double-peaked maximum, as 
observed in June 2014 by {\em XMM-Newton} \citep[][]{deMartino15}, is not 
present. The
modulation also appears different from those observed in Dec. 2013 and Jan. 
2014 by {\em XMM-Newton} and 
{\em Chandra}, respectively,  that instead are similar to each other 
\citep[see][]{Bogdanov14,deMartino15}\footnote{The
short {\em Chandra} observation is not used here}. 

\begin{figure}
\begin{center}
\includegraphics[width=2.0in,angle=-90, clip=true, trim=0 5 0 5]{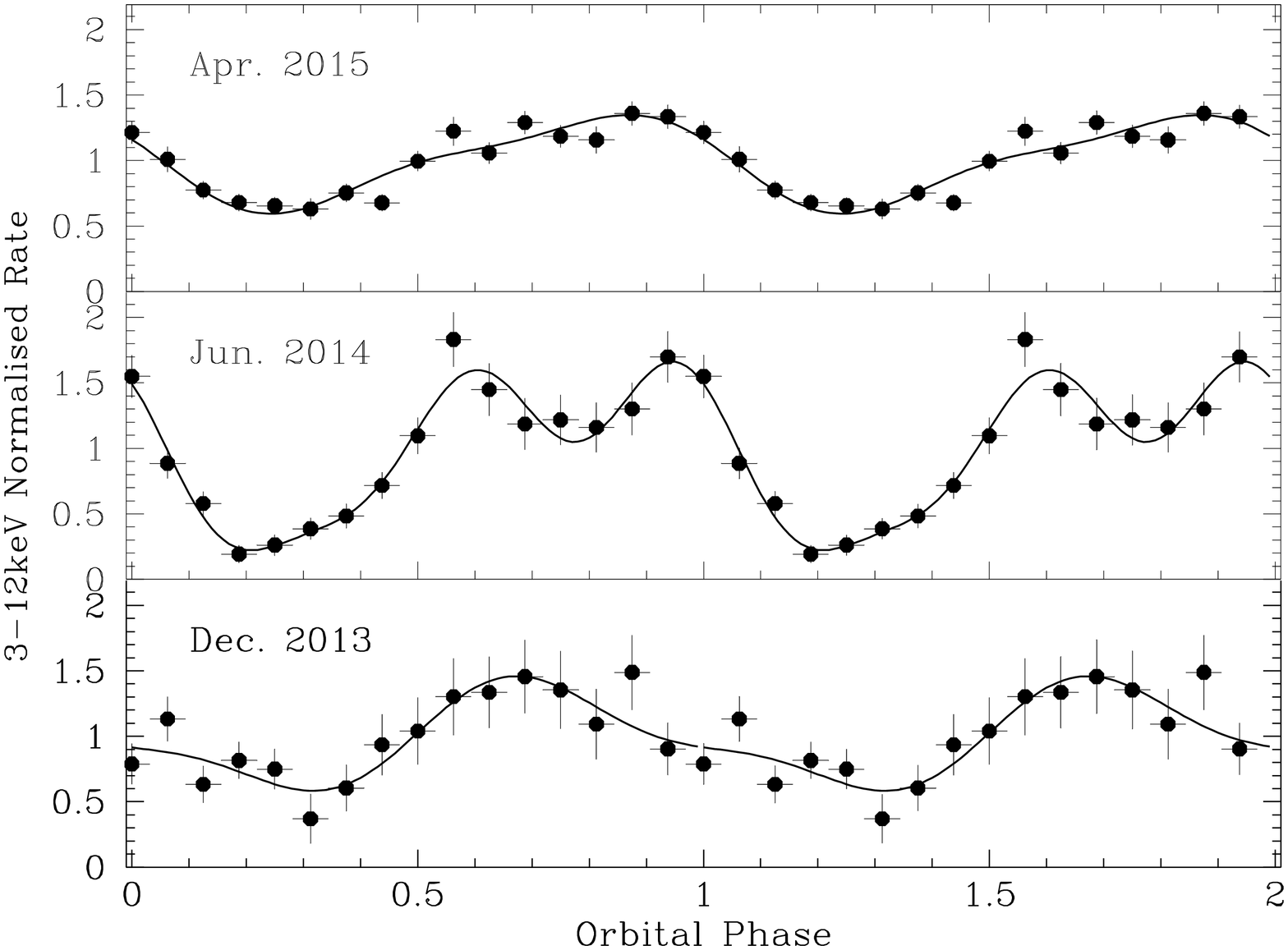}
\caption{Comparison of the background subtracted orbital folded light curves, 
evaluated over 16 phase bins, as observed by {\em XMM-Newton} in Dec. 2013 and
June 2014 and by   {\em NuSTAR} in Apr. 2015, in the common energy 
range 3-12\,keV. The light curves are normalised for a direct comparison.
Composite sinusoidal fits using  two (Dec. 2013 and Apr. 2015) or three (June 2014) 
frquencies, the fundamental $\rm f_b$ and two harmonics 2$\rm f_b$ and 
3$\rm f_b$, are also shown for each data set.}
\label{fold_lc_all}
\end{center}
\end{figure}

A close comparison of the 
{\em NuSTAR} data with  the two  {\em XMM-Newton} EPIC-MOS\footnote{Details 
of the reduction and extraction of the EPIC-MOS1,2 \citep{turner01} 
cameras are published in  \cite{deMartino15}}  
observations is made by folding background subtracted light curves 
in a common 3-12\,keV energy range using $\rm T_{asc}$ and $\rm P_b$ reported
in Table\,2. These are shown in Fig.\,\ref{fold_lc_all}. 
The orbital modulation is clearly structured in all data sets but shows 
differences especially in the broad maximum.
The fractional amplitude of the orbital modulation 
increases from $34(8)\%$ in Dec. 2013, to $67(9)\%$ in June 2014  
and decreases again to $36(2)\%$ in Apr.2015. Here we note that the
flux at minimum is about the same at all epochs, 
$\rm F_x \sim (2.0-2.5)\times10^{-13}\,erg\,cm^{-2}\,s^{-1}$. 
 While the variability 
in amplitudes between the first two observations was already reported in 
\cite{deMartino15}, the {\em NuSTAR} observation confirms that modulation 
amplitudes have not stabilized after the transition from a disc to a 
rotation-powered state, hinting to a non-stationary IBS  emission.
Noteworthy is the change in the shape of the maximum
among the three epochs,  displaying a local peak at $\rm \Phi_{orb}\sim$0.7 
in 2013, a double peaked maximum in 2014 centred at $\rm \Phi_{orb}\sim$0.75 
and a local peak at $\rm \Phi_{orb}\sim$0.9 in 2015. 
Since a simple sinusoid at the orbital frequency does
not satisfactory fit the light curves in the 3-12keV range, a fit was 
performed using 
a composite function consisting of two or three sinusoids at 
the fundamental frequency, $\rm f_b$, and its 
two higher harmonics, 2$\rm f_b$ and 3$\rm f_b$ (see Table\,\ref{sinu_fits}). 
Only for the 2014 data set
three components are required, while for the other two epochs two sinusoids
 at $\rm f_b$ and 2$\rm f_b$ well describe the orbital modulation, 
although the improvement over a single sinusoid is significant only 
at $\sim$90$\%$ and $92\%$ level for the 2013 and 2015 epochs, respectively.
 The amplitudes of the sine components at 
$\rm f_b$ and 2$\rm f_b$ are similar
in 2013 and 2015 within their uncertainties. An offset by 0.13 in 
phase is derived  in the  2$\rm f_b$ component between these two epochs
but not in the fundamental, that remains stable within errors. 
Different is the case observed in 2014 where $\rm f_b$ and 2$\rm f_b$ 
increase in amplitude by a factor of 1.7 and of 3, respectively. 
The  3$\rm f_b$ component 
has a fractional amplitude of $\sim 20\%$ and contributes at the minimum of the 
modulation.

\begin{table*}
\centering
\flushleft
 \begin{minipage}{140mm}
  \caption{\label{sinu_fits} Summary of sinusoidal fits to the X-ray folded 
orbital light curves in selected energy bands.}
  \begin{tabular}{@{}ccccccccccccc@{}}
  \hline
Year  & Band  & Flux$^a$ & \multicolumn{6}{c}{multi sinusoidal fit} & \multicolumn{3}{c}{single sinusoidal fit} \\
      & (keV) &     & $\rm A_{f_b}$ & $\rm \Phi_{f_b}$ & $\rm A_{2f_b}$ 
& $\rm \Phi_{2f_b}$ & $\rm A_{3f_b}$ & $\rm \Phi_{3f_b}$ & $\chi^{2}_{\nu}$/dof &  $\rm A_{f_b}$ & $\rm \Phi_{f_b}$ & $\chi^2_{\nu}$/dof \\

\hline 
 2013 & 3-12  &  3.8(4) & 0.37(7) & 0.52(3) & 0.13(6) & 0.48(4) & --      & --      & 0.89/11 & 0.34(8) & 0.51(9) & 1.30/13 \\
     & 0.3-3  & 1.4(1) & 0.22(5) & 0.50(1) & 0.11(5) & 0.38(7) & --      & --      & 1.00/11 & 0.25(5) & 0.49(1) & 1.53/13\\
     & 3-6    & 1.3(1) & 0.36(8) & 0.53(3) & 0.10(7) & 0.47(7) & --      & --      & 1.01/11 & 0.35(7) & 0.52(2) & 1.03/13 \\
     & 6-12 & 2.4(3)   & --      &  --     & --      & --      & --      & --      & --      & 0.4(2)  & 0.48(6) &  2.5/13 \\
& & & & & & & & & & & & \\
 2014 & 3-12   & 6.0(4) & 0.57(5) & 0.48(1) & 0.33(5) & 0.36(2) & 0.19(4) & 0.23(3) & 0.80/11 & 0.67(9) & 0.49(1) & 3.92/13 \\ 
     & 0.3-3  & 1.9(1) & 0.47(3) & 0.49(2) & 0.27(3) & 0.39(4) & 0.19(3) & 0.21(3) & 0.79/11 & 0.54(8) & 0.49(1) & 5.53/13 \\
     & 3-6    & 2.1(1) & 0.60(5) & 0.49(3) & 0.33(5) & 0.36(9) & 0.19(4) & 0.24(4) & 0.63/11 & 0.71(9) & 0.50(2) & 3.11/13 \\
     & 6-12   & 3.9(4) & 0.46(11)& 0.45(4) & 0.33(11)& 0.37(6) & 0.27(10)& 0.22(6) & 0.86/11 & 0.52(14) & 0.47(2) & 1.41/13\\
& & & & & & & & & & & & \\
2015 & 3-12 & 4.7(3) &  0.34(4) & 0.46(2) & 0.10(3) & 0.35(1) & --      & --      & 0.38/11 & 0.36(2) & 0.49(1) & 0.60/13\\
     & 3-6  & 1.7(2) &  0.28(5) & 0.45(3) & 0.14(5) & 0.32(6) & --      & --      & 1.33/11 & 0.29(4) & 0.46(2) & 1.87/13 \\
     & 6-12 & 3.1(2) &  --       & --      &  --     & --    & --      & --      & --      & 0.39(3) & 0.47(1) & 1.42/13 \\
     & 12-25 & 5.8(2) & 0.36(2) & 0.44(1) & 0.14(2) & 0.36(1) &  --    & --     & 0.54/11 &  0.34(3) & 0.44(1) & 1.05/13 \\
     & 25-79 & 19.8(1) & --     &  --     & --    &  --     &   --   & --    &  --   & --  & -- & --   \\

\hline
\end{tabular}
$^a$ Average unabsorbed fluxes in units of $\rm 10^{-13}\,erg\,cm^{-2}\,s^{-1}$ in selected energy bands 
   as derived from spectral fits to the average spectra at each epoch.
\end{minipage}
\end{table*}

\begin{figure*}
\begin{center}
\includegraphics[width=2.0in,angle=-90, clip=true, trim=0 5 0 5]{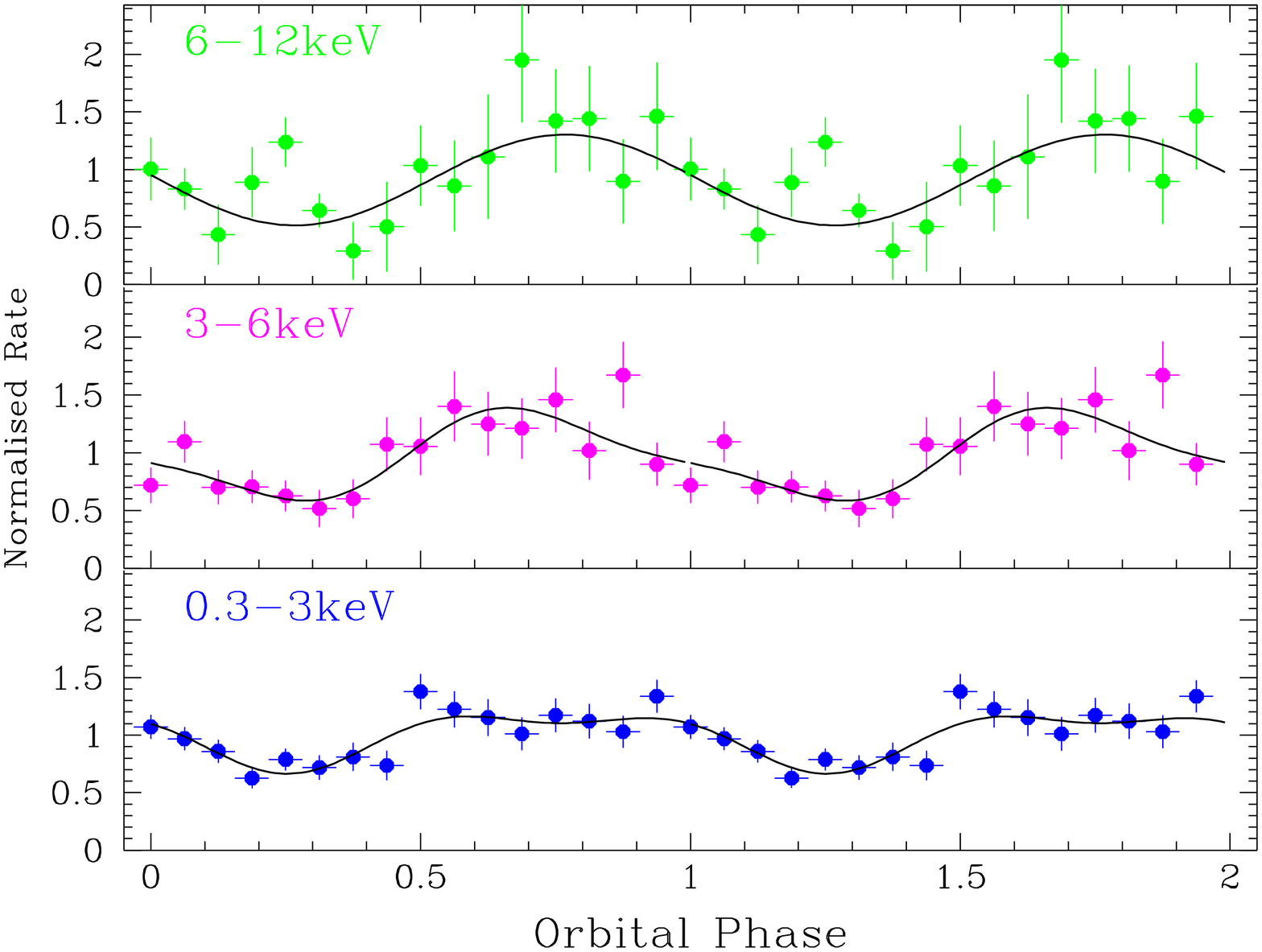}
\includegraphics[width=2.0in,angle=-90, clip=true, trim=0 5 0 5]{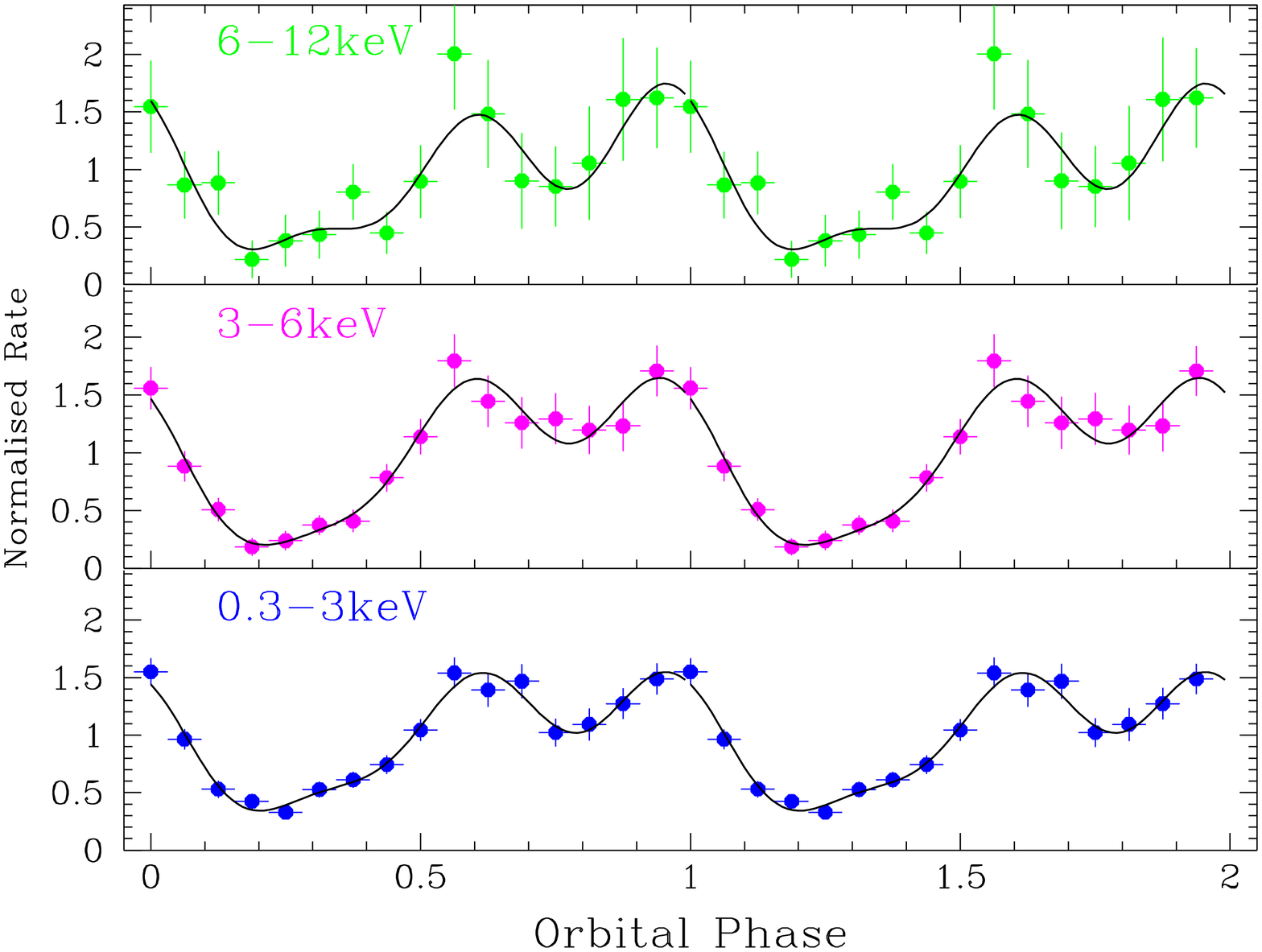}
\caption{ Energy resolved light curves folded at the 6.91\,h orbital period 
in selected
bands evaluated over 16 phase bins, as observed with {\em XMM-Newton} in Dec. 2013
({\it left panel}) and in Jun.2014 ({\it right panel}).
For the former epoch, a composite sinusoidal fit with two components is 
also shown 
for the softer bands while only the fundamental is shown for the harder 
6-12\,keV range.
Instead, three sinusoids are required for the second observation. 
Ordinates are normalised to unity.}
\label{fold_2013_2014}
\end{center}

\end{figure*}

A further comparison is carried out using common selected 
energy ranges between the
{\em XMM-Newton} and {\em NuSTAR} data,   namely
3-6\,keV, 6-12\,keV for the three epochs. Folded light curves
in the soft 0.3-3\,keV bands were also produced for the
two {\em XMM-Newton} data sets to inspect the energy dependence of the
orbital variability. Table\,\ref{sinu_fits} reports the results of the fits
by using up to three components.
In Fig.\,\ref{fold_2013_2014} the energy resolved light curves 
in the three energy ranges are shown for the two earlier epochs.
Two sine components  better describe the orbital modulation 
in the softer  0.3-3\,keV and 3-6\,keV bands in Dec. 2013, 
although the improvement over a single sinusoid is signficant only 
at $\sim 96\%$ and at $\sim64\%$, respectively. At higher (6-12\,keV) 
energies the variability is badly defined in 2013, though consistent with
that observed by {\em NuSTAR} in 2015.

\begin{figure*}
\begin{center}
\includegraphics[width=2.0in,angle=-90, clip=true, trim=0 5 0 5]{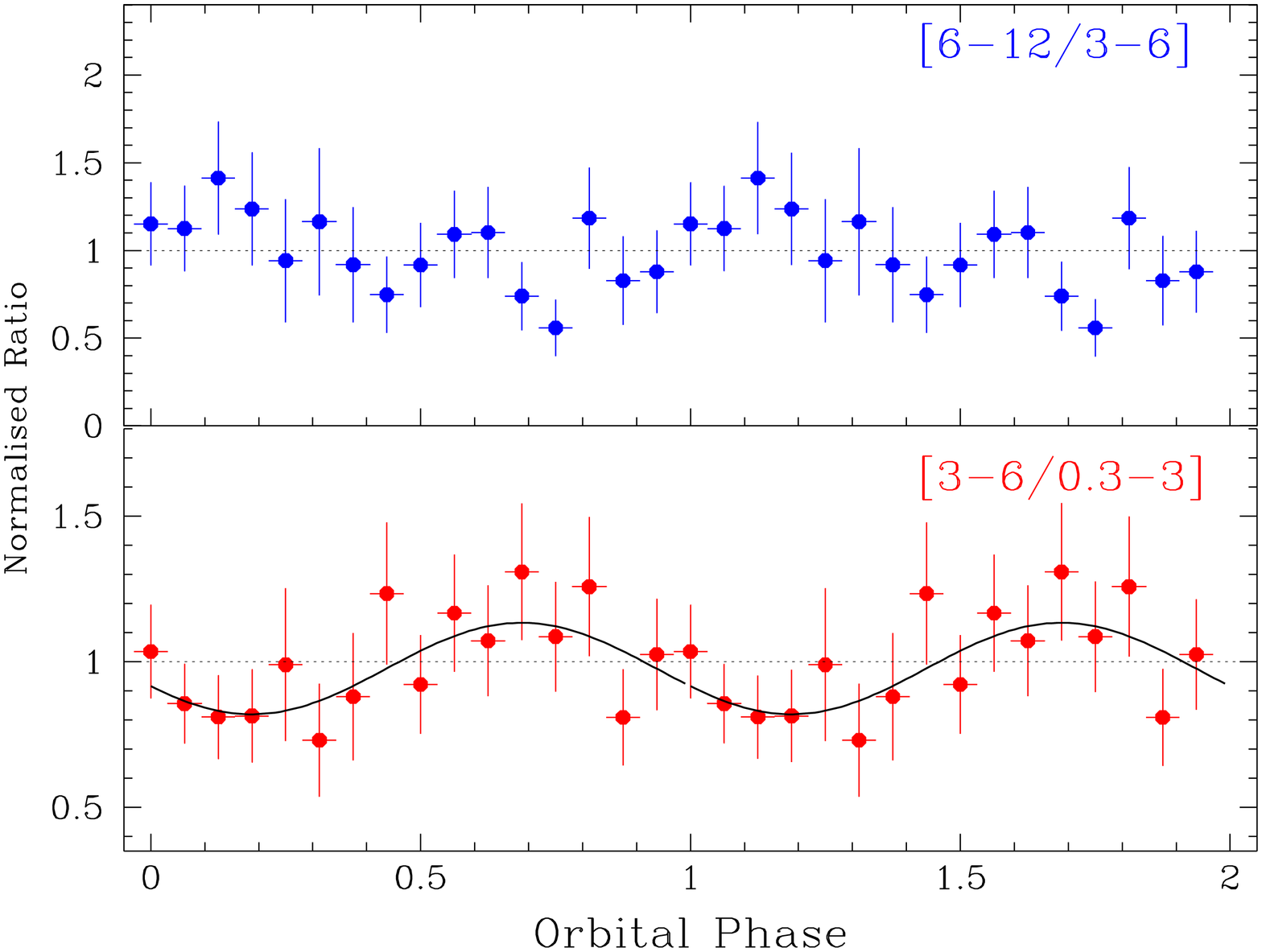}
\includegraphics[width=2.0in,angle=-90, clip=true, trim=0 5 0 5]{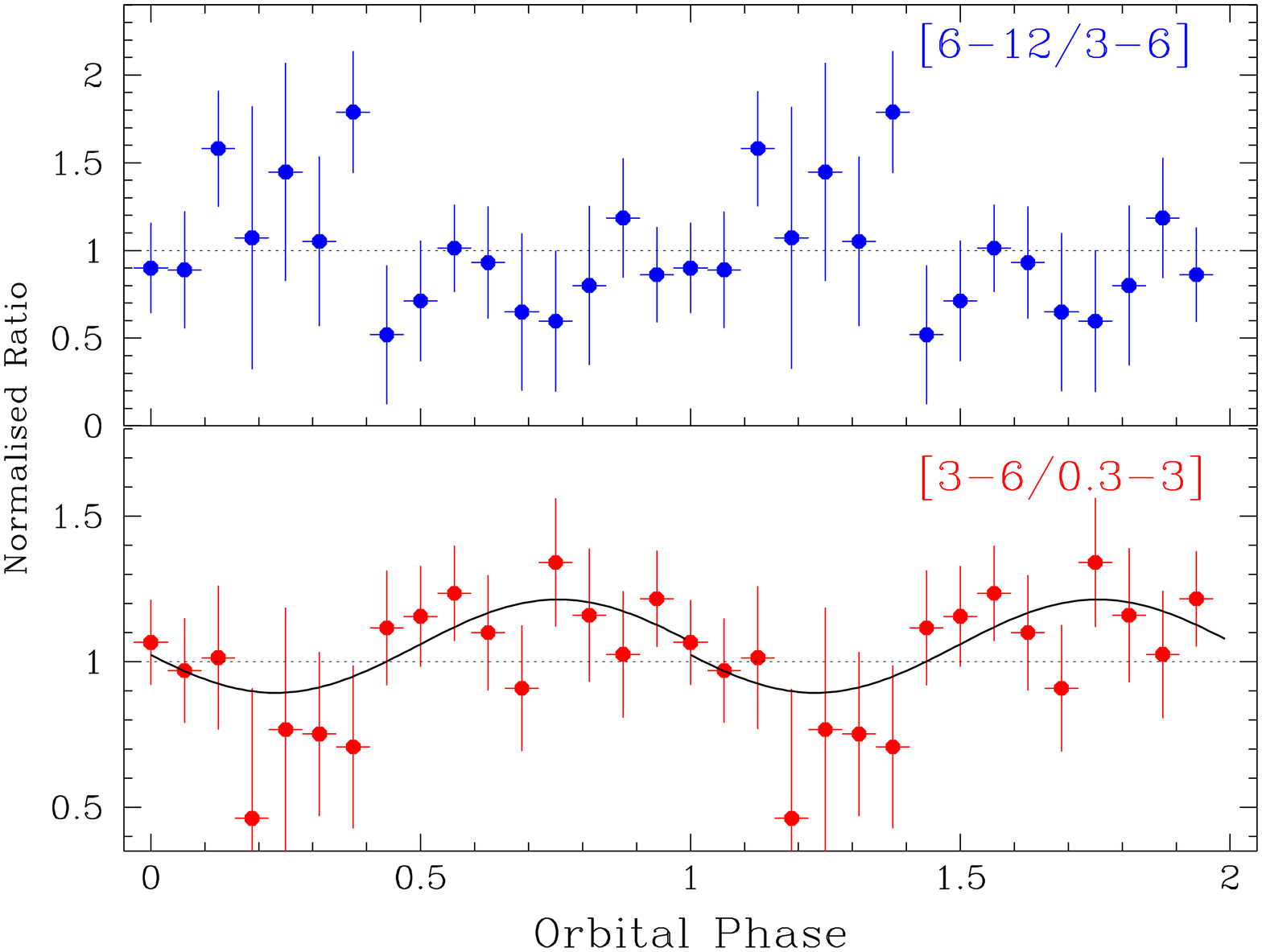}
\caption{Hardness ratios folded at the 6.91\,h orbital period in the 0.3-3\,keV, 
3-6\,keV and 6-12\,keV 
ranges evaluated over 16 phase bins as observed by {\em XMM-Newton} in Dec. 2013
({\it left}) and in Jun.2014 ({\it right}). A sinusoidal function (black) is
also shown to help visualization for the mild hardening between the 
soft 0.3-3\,keV and medium 3-6\,keV ranges (bottom panels) although 
significance of the variability is below 3$\sigma$ (see text for details).
Ordinates are normalised to unity.}
\label{HR_2013_2014}
\end{center}
\end{figure*}

\noindent As for Jun. 2014,  three 
sinusoids are required in all bands.
As before, the amplitudes of the fundamental $\rm f_b$ frequency change by 
a factor of about two while that of 2$\rm f_b$  
increases by $\sim$3 in all bands with respect to the 2013 and 2015 observations.
 The  modulation is stronger in the 3-6\,keV range, while 
in the  6-12\,keV band, it is badly defined and 
consistent with the softer ranges within uncertainties.  This produces a mild
hardening between 0.3-3\,keV and 3-6\,keV ranges at 
$\rm \Phi_{orb}\sim 0.75$, i.e. at the inferior conjunction of the NS.
 This behaviour is seen in the hardness ratios (HR), defined as the 
the ratio of count rates in the 3-6\,keV and 0.3-3\,keV bands and
6-12\,keV and 3-6\,keV ranges.  At higher energies 
the HR are instead dominated by noise (see Fig.\,\ref{HR_2013_2014}). 
The significance of the orbital 
variability of the hardness ratios (HR), 
defined as the ratio of count rates in the 3-6\,keV and 0.3-3\,keV bands and
6-12\,keV and 3-6\,keV ranges, was inspected by modelling the orbital HR 
curves with a constant and a sinusoidal function. For the June 2014 observation,
and for the softer HR between 3-6\,keV and 0.3-3\,keV we obtained 
$\chi^2$=10.7, d.o.f.=15 in case of a constant 
and $\chi^2$=6.5, d.o.f.=13 
for a sinusoidal fit with fractional amplitude of 17(6)$\%$.
The improvement of the modelling with respect to a
constant was however at only 2$\sigma$ level. For the Dec. 2013 observations 
and for the
same soft HR we obtained in the case of a costant $\chi^2$=12.95, d.o.f.=15 
and $\chi^2$=6.3, d.o.f.=13 for a fit with a sinusoid with fractional amplitude
of $15(5)\%$. The variability is significant at only 2.6$\sigma$ level. 
The HR at higher energies between 3-6\,keV and 6-12\,keV are instead 
consistent with a constant in both
2013 and 2014 observations (see Fig.\,\ref{HR_2013_2014}). 
This refines the results found by \cite{deMartino15}, where hardness ratios
were inspected in 0.3-2\,keV and 2-10\,keV.

\begin{figure*}
\includegraphics[width=2.5in,height=3.in,angle=-90, clip=true, trim=0 5 0 5]{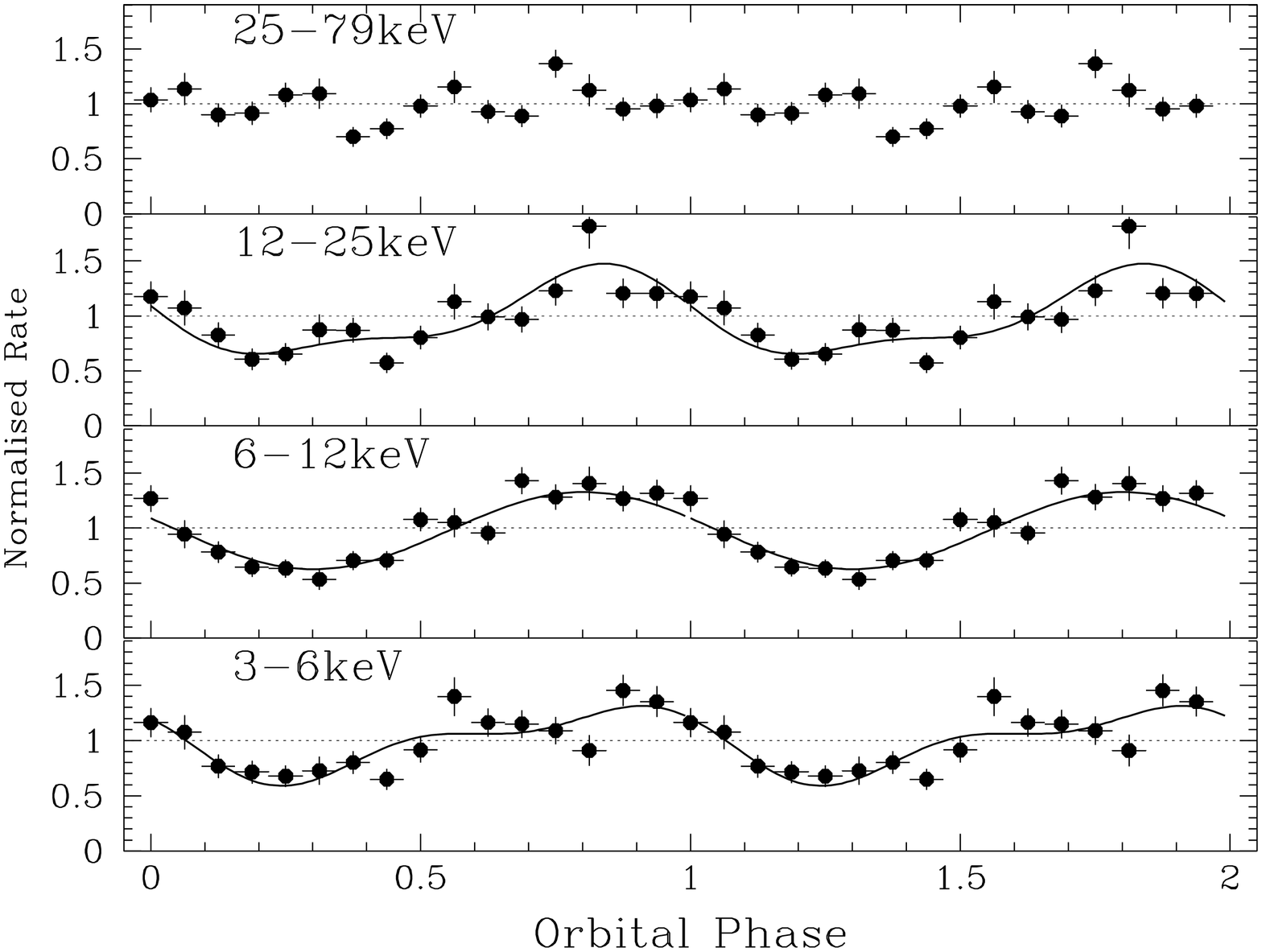}
\includegraphics[width=2.5in,height=3.in,angle=-90, clip=true, trim=0 5 0 5]{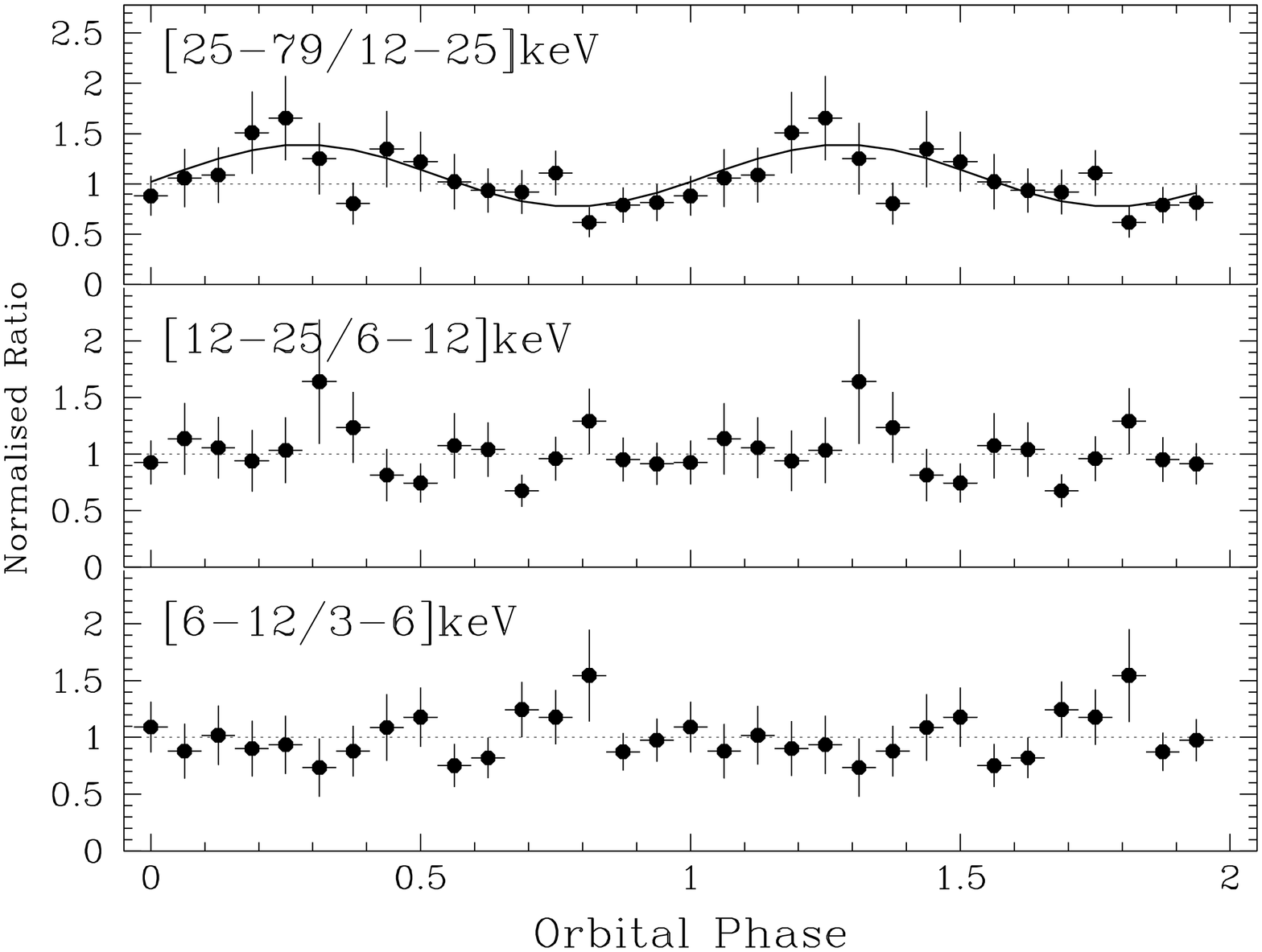}
\caption{{\it Left:} Folded orbital light curves  in selected energy bands 
and evaluated in 16 phase bins as observed by {\em NuSTAR} 
in Apr.2015. A composite sinusoidal fit with one or two harmonics 
(see Table\,\ref{sinu_fits}) is also 
shown, except for the harder band 25-79\,keV where no modulation is detected.
{\it Right:} Hardness ratios remain constant except in the hardest range due
to the lack of detection of orbital variability. Ordinates are normalised to unity.}
\label{nustar_lc_hr}
\end{figure*}

The {\it NuSTAR} orbital light curves of Apr. 2015 were also produced in 
the 12-25\,keV, 25-40\,keV, 40-79\,keV and 25-79\,keV 
ranges (see  Fig.\,\ref{nustar_lc_hr}). A weaker 
modulation than that observed  in Jun.2014, but of similar 
amplitude as in Dec. 2013 ($\sim30\%$), is detected below 12\,keV. 
The  modulation  extends up to $\sim$25\,keV with similar fractional amplitudes. 
It is however undetected at higher energies with an 
upper limit to the modulation amplitude of $\sim$10$\%$ (3$\sigma$ confidence 
level). Two harmonics account for the  asymmetric shape in the 
3-6\,keV and 12-25\,keV range
although significant at $92\%$ and $97\%$ confidence levels, respectively. 
In the 6-12\,keV range only
the fundamental frequency is required (see Table\,\ref{sinu_fits}). 
The hardness ratios in the four selected bands reveal no 
spectral variations except at higher energies due to the lack of 
detectable modulation above 25\,keV (right panel in Fig.\,\ref{nustar_lc_hr}). 

We also performed a Fourier search for coherent periodicities in the 3-79\,keV 
time-series recorded by both {\em NuSTAR} modules, after preliminarily
correcting the photons times of arrival for the orbital motion of the
pulsar using the orbital parameters listed in Table\,3. The {\em NuSTAR} onboard 
 clock suffers of 
timing jumps introducing spurious derivatives of the order of
$10^{-10}$~Hz~s$^{-1}$ of the frequency of a $\sim 600$~Hz coherent
signal \citep[see][]{Sanna17}.
Therefore a search was done over a frequency range of $+/- 10^{-4}$~Hz around the spin
frequency of J1227. The maximum value of the Fourier power density
observed corresponds to an upper limit on the rms pulse amplitude of
$8.8\%$ (3$\sigma$ confidence level; \citep[see][]{Vgh94}).
This value is comparable with the upper limit set from
the analysis of {\em XMM-Newton}  observation performed in 2014 (7.1\%;
\citep{Papitto15}
and slightly smaller than the
amplitude of $11\pm2\%$ of the  pulses detected in the 0.25-2.5 keV
band during the radio pulsar state of PSR\,J1023+0038 \citep{Archibald10}.

\subsection{The optical variability }

\begin{figure}
\includegraphics[width=2.4in,angle=-90, clip=true]{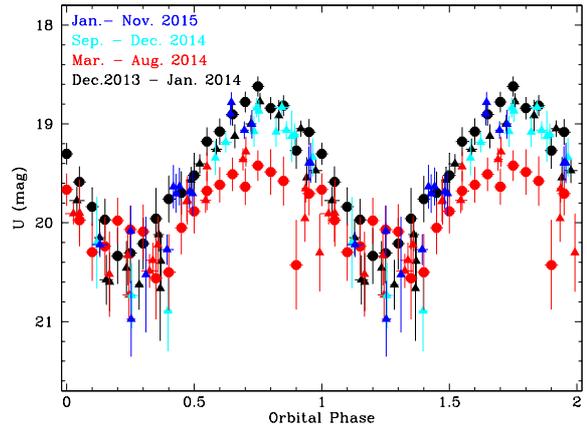}
\caption{The U-band folded light curves of J1227 at the orbital period 
acquired by {\em Swift} UVOT
from Dec. 2013 to Nov. 2015 (triangles) color coded as 
Dec.2013-Jan. 2014 (black), Mar.-Aug. 2014 (red), Sept.-Dec. 2014 (cyan) and
Jan.-Nov. 2015 (blue). The {\em XMM-Newton} OM (filled circles) light curves
in Dec. 2013 (black) and Jun. 2014 (red) \citep{deMartino15} 
are also reported for comparison.
J1227 displays a lower amplitude modulation by a factor $\sim$2 as
observed by both instruments only in the Mar.-Aug.2014 period.
}
\label{opt_fold}
\end{figure}

The optical long-term behaviour of J1227 was also inspected by 
extending the comparison of the
orbital modulations  between the X-ray and optical ranges at the epoch of the 
{\em NuSTAR} observation. 
In particular, the optical U-band modulation decreased by a factor of $\sim$2
between the {\em XMM-Newton} pointings in Dec.\,2013 and Jun.\,2014 and 
accompained by an anti-correlation in the X-ray band \citep{deMartino15}.
Therefore U-band photometry was extracted using {\em Swift} UVOT \citep{roming05} 
observations acquired between Jan. and Nov. 2015.This
larger time-span was dictated by the occasional coverage with the U filter.
As an additional comparison U-band measures were also extracted
for several epochs, namely Dec.\,2013-Jan.\,2014,
Mar.-Aug.\,2014 and Sept.-Nov. 2014 to trace potential changes in the optical
light curve over different epochs. 
For this purpose, we used the tool {\sc uvotmaghist} in {\sc ftools} 
v.6.22, accounting for sensitivity UVOT patches\footnote{Small scale 
sensitivity patches in the UVOT detector are described at 
\url{https://swift.gsfc.nasa.gov/analysis/uvot_digest/sss_check.html}}, 
to extract the photometry 
and applied the correction to the Solar System  barycentre.
The folded light curves were then compared
with those observed in 2013 and 2014 by the Optical Monitor (OM) on-board 
{\em XMM-Newton}. The modulation observed during the Jan.-Nov.\,2015 
has a peak-to-peak amplitude of 1.47(8)\,mag, similar to that observed in 
Dec. 2013 by the OM (1.43(4)\,mag) and to those observed
by UVOT in the same period (Dec.\,2013-Jan.\,2014) and in Sept.-Nov.2014.
It is thus larger by a factor of $\sim$2 than that observed in Jun.\,2014 
by the OM (0.72(9)\,mag)) and by UVOT 
in the Mar.-Aug.\,2014 (see Fig.\,\ref{opt_fold}).  
This  indicates that
the smaller orbital U-band modulation amplitude was only reached during 
the Mar.-Aug.\,2014 period,
whilst before and after that epoch the variability was larger. 
Here we also confirm that J1227 shows within uncertainties of the photometry
the same minimum optical flux  at all epochs ($\rm U \sim 20.5$).
The decrease and 
the subsequent increase in amplitude in 2014 thus occurred over a 
timescale of $\sim$2 and 1\,month, respectively.
Comparing the optical and X-ray behaviour in the 2013, 2014 and 2015 
epochs, we can firmly confirm the the orbital modulation 
is  anti-correlated in the two energy regimes, i.e. 
when the X-ray modulation is larger, the optical U-band  orbital 
variability is smaller.

\section{Spectral analysis}

The average 3-79\,keV spectra from the {\em NuSTAR} FPMA and FPMB modules 
over the whole observation are featureless and of similar shape as those 
previously observed in the softer 0.3-10\,keV range by 
\cite{Bogdanov14,deMartino15}. 
The spectrum extends up to 70\,keV without any apparent spectral break. 
An absorbed power law model was adopted to fit the spectra with a multiplicative 
normalization constant to account for the differences of the FPMA and 
FPMB detectors, namely {\sc const*tbabs*powerlaw} in 
{\sc xspec}\footnote{Details on {\sc xspec} models available at 
\url{https://heasarc.gsfc.nasa.gov/xanadu/xspec/manual/Models.html}}.
However, since the the hydrogen column density  
found from the {\em XMM-Newton} spectra
in 2013 and 2014 \citep{Bogdanov14,deMartino15} were both agreeing with an upper
limit of  $\rm N_H <  5\times10^{20}\,cm^{-2}$, and due to the
lack of sensitivity of the {\em NuSTAR} data to such low values, 
the hydrogen column density was fixed to zero.
We then find  a power law photon index $\Gamma=1.17\pm0.08$ 
(90$\%$ confidence level) and an unabsorbed flux of 
$\rm 3.4\pm0.1 \times 10^{-12}\,erg\,cm^{-2}\,s^{-1}$ in the 
3-79\,keV range  (see Table\,\ref{spec_fit}).
The $\Gamma$-index is fully consistent with 
that we derived in the {\em XMM-Newton} data of Dec.\,2013 
(1.07$\pm$0.08) and Jun.\,2014 (1.2$\pm$0.1). If a power law with an exponential 
cutoff is instead used, a worse fit is obtained ($\chi^2_{red}$=1.13, d.o.f.=136) 
with a power law photon index essentially 
the same, $\Gamma=1.15\pm0.08$, and a cutoff energy $\rm >$170\,keV.  
The average  flux in the 3-12\,keV
band is $\rm 4.7\pm0.3 \times 10^{-13}\,erg\,cm^{-2}\,s^{-1}$, similar to that
observed in Dec. 2013 but lower than in Jun.\,2014. 
With the above spectral
parameters the  X-ray luminosity in the 3-79\,keV is 
$\rm 7.6_{-0.8}^{+3.8} \times 
10^{32}\,erg\,s^{-1}$, for the adopted  distance of 1.37$_{-0.15}^{+0.69}$\,kpc,
accounting for the distance uncertainties.

\noindent To extend the spectral analysis into the soft energy range, 
{\em Swift} XRT \citep{burrows05} data were then used. The monitoring 
of J1227 over the years encompasses observations performed
in 2015, including April 25 (ObsId:81457001, 1.97\,ks), but the short 
coverage results in a 0.3-10\,keV spectrum with poor statistics. 
We then extracted XRT PC mode observations performed over a period of 4 
months and with exposures longer than 1\,ks, namely March 11, 2015 
(ObsId:35101021, 2.78\,ks), April 25, May 3, 2015 (ObsId:81457002 - 1.93\,ks), 
June 22, 2015 (ObsId: 35101024, 2.22\,ks) and July 3, 2015 (ObsId: 35101025), 
totalling 9.9\,ks. The XRT PC mode data were extracted  using the automatic 
analysis software \citep{Evans09} from the UK {\em Swift} Science Centre.    
During this period J1227 was found
at an average count rate level (PC mode) of 
$\sim \rm 5.7\times10^{-3}$\,cts\,s$^{-1}$.
The XRT 0.3-10\,keV background subtracted light curve folded at the 6.91\,h orbital period
although with small gaps displays a minimum and a maximum at consistent phases 
as those observed in
{\em XMM-Newton} and {\em NuSTAR} data but the fractional 
amplitude of the variability is 
poorly constrained (46$\pm$23$\%$), preventing a meaningful comparison. 
The accumulated XRT 0.3-10\,keV spectrum grouped to 
have 5 counts per bin, 
was fitted with an absorbed power law, giving a photon index
$\Gamma$=1.1$_{-0.4}^{+0.8}$, an upper limit to the column 
density $\rm N_H < 2.4\times10^{21}\,cm^{-2}$ and an unabsorbed flux of 
$\rm 3.6_{-0.4}^{+0.9}\times10^{-13}\,erg\,cm^{-2}\,s^{-1}$.   
When evaluated in the 3-6\,keV range, 
the fluxes between {\em NuSTAR} and  {\em Swift} agree within errors despite
the uneven orbital coverage of the XRT instrument. 

\begin{figure}
\includegraphics[width=2.3in,angle=-90, clip=true]{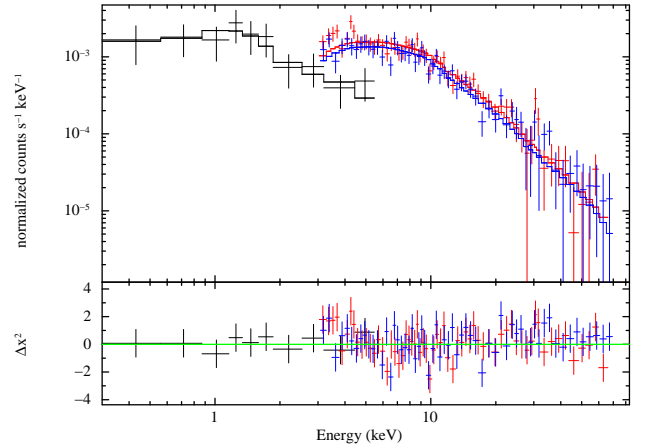}
\caption{The average Mar.-Jul.\,2015 {\em Swift} XRT (black) and Apr.\,2015 
{\em NuSTAR} FPMA (red) and 
FPMB (blue) spectra fitted with an absorbed power law model
with $\Gamma$=1.17$\pm$0.08, fixing $\rm N_H=5\times10^{20}\,cm^{-2}$.}
\label{spec_ave}
\end{figure}

\noindent We then combined {\em Swift} and {\em NuSTAR}  average spectra to perform
a broad-band spectral fit (Fig.\,\ref{spec_ave}). Using the same model as 
before, leaving
free the hydrogen column density of the absorber we obtain
$\rm N_H < 1.2\times10^{21}\,cm^{-2}$ ($\chi^2_{\nu}$=1.0, d.o.f.=126) 
and all other parameters
consistent with those without the addition of the {\em Swift} spectrum
(Table\,\ref{spec_fit}). 
We then adopted a fixed 
hydrogen column density to the value found with the {\em XMM-Newton} data. 
The total 0.3-79\,keV unabsorbed flux is 
$\rm 3.7_{-0.4}^{+0.2}\times10^{-12}\,erg\,cm^{-2}\,s^{-1}$. This gives 
a luminosity of $\rm 8.3_{-0.9}^{+4.2}\times10^{32}\,erg\,s^{-1}$ in the
same energy range for D=1.37$_{-0.15}^{+0.69}$\,kpc.

\begin{table*}
\begin{center}
  \caption{\label{spec_fit} Spectral fit parameters of the {\em NuSTAR} spectra. Uncertainties are
at the 90$\%$ confidence level. The absorbing column density 
was set to 0 (see text for details).}
  \begin{tabular}{@{}cccccc@{}}
  \hline
Parameter            & Average & \multicolumn{4}{c}{Orbital phase} \\
                     &         & $\rm \Delta \Phi_{min}$  & $\rm \Delta \Phi_{max}$ & $\rm \Delta \Phi_{rise}$ & $\rm \Delta \Phi_{decay}$ \\

                     &         &   0.17-0.45  &   0.69-0.99  &  0.45-0.69 & 0.00-0.17  \\
\hline
$\rm C_{FPMB}^{*}$  &  0.92$\pm$0.07 &  0.88$_{-0.25}^{+0.26}$ & 0.95$_{-0.11}^{+0.13}$         & 0.80$_{-0.13}^{+0.15}$ & 0.94$_{-0.19}^{+0.23}$  \\
$\rm \Gamma$       &  1.17$\pm$0.08 & 1.13$_{-0.21}^{+0.27}$  & 1.15$\pm$0.13& 1.26$_{-0.17}^{+0.18}$ & 1.23$_{-0.24}^{+0.25}$ \\
Flux$^{**}$        &  3.40$\pm$0.10 & 1.96$_{-0.59}^{+0.81}$  & 5.05$_{-0.81}^{+0.93}$ &  3.62$_{-0.76}^{+0.92}$ & 2.94$_{-0.84}^{+1.12}$ \\
$\chi^2_{\nu}$/d.o.f. &  1.05/119   & 1.48/19                 & 1.17/33       & 0.76/20 & 0.76/20 \\
\hline
\end{tabular}
\flushleft
$^{*}$ Scaling constant for FPMB spectrum\\
$^{**}$ Unabsorbed flux in units of $\rm 10^{-12}\,erg\,cm^{-2}\,s^{-1}$ in the 3-79\,keV range\\  
\end{center}
\end{table*}

\begin{figure}
\includegraphics[width=2.3in,angle=-90, clip=true]{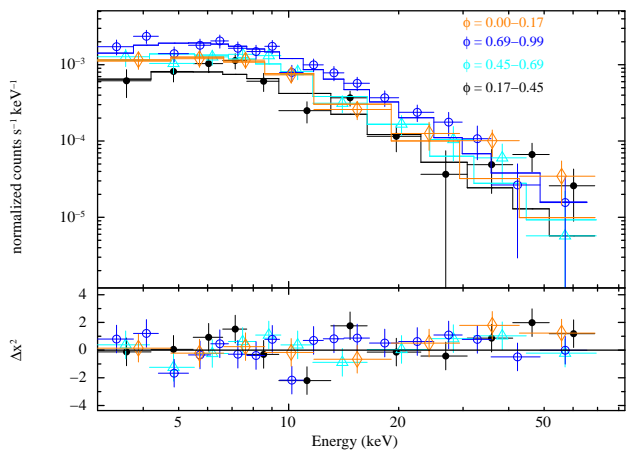}
\caption{The phase-resolved {\em NuSTAR} spectra at orbital minimum (black 
filled points), 
maximum (blue empty circles), at rise to maximum (light blue empty
triangles) and at decay to minimum (orange empty diamonds)
are shown together with their best spectral model parameters reported 
in Table\,\ref{spec_fit}.
For clarity purposes only the FPMB spectra are reported.}
\label{spec_resolved}
\end{figure}

Hints of changes in the spectral shape in the {\em XMM-Newton} data 
of Dec.\,2013 and Jun.\,2014 
were found between the orbital maximum and minimum, with a tendency of a 
harder power-law index at maximum  ($\Gamma$=1.1$\pm$0.1) with respect
to that at  minimum ($\Gamma$=1.30$_{-0.15}^{+0.25}$) 
\citep[see][]{deMartino15}. As seen in the  hardness ratios  in the
selected energy ranges (Fig.\,\ref{HR_2013_2014}) a tendency of a lower amplitude
in the softer 0.3-3\,keV band was found, but not at higher energies 
(see Sect.\,4.1).  This could produce a hardening of the 
spectrum at orbital maximum.  
The energy resolved {\em NuSTAR} orbital light curves also confirm the 
lack of energy
dependence of the amplitudes above 3\,keV and that the modulation 
is undetected above 25\,keV.
We however inspected the {\em NuSTAR} spectra along the orbital period 
by selecting  four orbital
phase ranges, namely the maximum ($\Delta \Phi_{orb}$=0.69-0.99), the minimum 
($\Delta \Phi_{orb}$=0.17-0.45), the rise to maximum ($\Delta \Phi_{orb}$=0.45-0.69) 
and the decay to the minimum ($\Delta \Phi_{orb}$=0.0-0.17) 
(see Table.\,\ref{spec_fit}). 
The variations essentially occur in the normalization with a power law photon index being
constant within uncertainties, although there is a tendency of a softer 
spectrum in the rise to and in the decay from maximum.
The lack of orbital modulation above $\sim$25\,keV is also apparent
in the phase-resolved spectra (Fig.\,\ref{spec_resolved}).

\section{Discussion}

We here discuss the first X-ray data above 10\,keV obtained with {\em NuSTAR} 
of the tMSP J1227  during the rotation-powered state complemented 
with a 2.5\,yr-long radio coverage at {\em Parkes} telescope.  

\subsection{The radio monitoring}

The  radio monitoring spanning from June 2014 to February 2017 and 
carried mainly at 20-cm (1.4\,GHz) provided us with a pulsar spin frequency 
at higher accuracy than previously reported \citep{Roy15} and a new 
orbital ephemeris,
which is refined close to the April 2015 {\em NuSTAR} observation.
The radio eclipses at 20-cm extend at least from 
$\rm \Phi_{orb}\sim$ 0.06 to 0.39, although at some epochs the pulsar 
was not detected for a full orbital cycle. This gives a lower limit to the eclipse
extent $\rm \Delta \Phi_{orb}\sim 0.3$. However at lower frequencies 
(e.g. 50-cm), the eclipse length could be as long as 
$\rm \Delta \Phi_{orb}\sim$0.52-0.56 (Figure\,\ref{fig:radio_ecli}). 
As a comparison, \cite{Roy15}
derive an eclipse length of $\rm \Delta \Phi_{orb}\sim 0.40$ at 607\,MHz.
Long eclipses up to $60\%$ of 
the orbital cycle  are typical of RBs  rather 
than BWs that instead have shorter eclipses ($\sim 5\%-15\%$).
In both types of systems the
eclipses are a signature of intrabinary material,  
produced by the interaction of the pulsar and companion 
star winds, likely in the form of a shock.

Using the thin-shell approximation for the collision of two 
isotropic winds \citep{canto96} and, for simplicity, that the shock axis of symmetry
lies in the 
orbital plane \citep{Romani16,Wadiasingh17}, the maximum shock opening angle 
as measured from the pulsar is related
to the binary inclination and eclipse length  through a simple trigonometric relation:
$\rm cos\,\theta_{max} = sin\,i\,cos(\pi\,\Delta\Phi_{orb})$. For J1227, 
assuming a very conservative lower limit to the radio eclipse length  
$\rm \Delta \Phi_{orb}\sim$0.4 and a binary inclination 
$\rm 46^o \lesssim i \lesssim 55^o$
\citep[][]{deMartino15,Rivera_Sandoval18}, the maximum
shock opening angle would correspond to $75-77^o$ or
$\sim (5/12)\,\pi$. 
Following \cite[][]{canto96}, the asymptotic shock angle
is given by: $\rm \theta_{\infty} - tan\,\theta_{\infty} = 
\pi / (1-\eta^{-1})$, where  $\eta = \rm \dot M\,v_w\,c/\dot E$
is the ratio of the secondary star to the pulsar wind momentum. 
The latter is related to the shock stagnation point radius, here from the MSP,  
as: 
$\rm R_s/a = \eta^{-1/2}/(1+ \eta^{-1/2})$, where $\rm a$ is the binary
separation \citep[][]{canto96}. 
The estimated maximum shock opening angle would then give 
a maximum distance of the shock from the pulsar $\rm (R_s/a)_{max} \sim 0.60$
and $\eta\sim 0.43$.   
Adopting the projected semi-major 
axis of the pulsar, $\rm a1$ = 0.668492\,lt-s
(Table\,\ref{tab:ephemTOT}), the above binary inclination range and
a mass ratio  $\rm q\sim$ 0.194 \citep{Roy15}, we obtain an orbital separation
$a\sim  1.5-1.7\times10^{11}$\,cm.  Consequently we estimate
$\rm R_{s,max} \sim 0.6\,a \sim 1.0\times10^{11}\,cm$, which is about the
NS Roche Lobe radius. However, in order to have the IBS formed around the pulsar, 
as indicated by the phasing 
of the X-ray light curves (see below),  the condition 
$\rm (R_s/a)_{max} \lesssim 0.50$ should be satisfied. The eclipse fraction
observed at high frequencies then cannot provide a reliable estimate, but only a 
very loose lower limit to the eclipse fraction.  In fact,  
RBs have longer eclipses at lower frequencies, with a 
dependence as $\rm \sim \nu^{-0.4}$ \citep{Broderick16}.  
As a comparison, for PSR\,J1023+0038 \cite{Wadiasingh17} estimate 
$\rm (R_s/a)_{max}  \lesssim 0.4$ for an eclipse fraction $\gtrsim 0.6$ at 350\,MHz \citep{Archibald13}. 
Our monitoring at
{\em Parkes} encompasses only a few observations  at 728\,MHz and the 
previous measures by \cite{Roy15} also had a few at 322\,MHz. 
Scaling the eclipse length from 607\,MHz to 322\,MHz, we derive 
$\rm \Delta \Phi_{orb} \sim 0.52$. Using instead this more likely value 
for the eclipse fraction, we obtain $\rm (R_s/a)_{max} \sim 0.45$ 
and $\eta\sim 1.5$.
As noted by \cite[][]{Wadiasingh17} caution should be taken in the direct 
interpretation of $\eta$ in terms of wind ram pressure 
due to the gravitational influence of the pulsar. 
Additionally, the pulsar wind is likely to be more concentrated 
equatorially
and an additional dependence of the MSP wind momentum as $\rm sin^n \theta_{\star}$
with $\rm n \sim 2$, or even 4 for an oblique rotator, introduces changes in the
shape of the IBS narrowing it at the poles \citep{Romani16,Kandel19}.

\subsection{The X-ray emission} 

The X-ray spectral shape is a power law with index of 1.2 with no spectral 
break up to $\sim$70\,keV, indicative of synchrotron cooling. The large
orbital variability observed in J1227 and in other RBs, suggests that
the IBS dominates the X-ray emission. This emission depends on the 
post-shock magnetic field strength ($\rm B_2$) and on  the ratio 
of magnetic to particle energy density
$\sigma$  \citep[see][]{Kennel_Coroniti84}.
In the case of weakly magnetised winds, $\sigma$ could be as low 
as $\sim 3\times10^{-3}$ as in the Crab,
while in strongly magnetised winds $\sigma >> 1$. 
The post-shock magnetic field is related to the upstream magnetic field $\rm B_1$
as $\rm B_2 \sim 3\,B_1$ and  $\rm B_2 \sim B_1$ in 
these extreme cases, respectively \citep{Kennel_Coroniti84}. 
At large distances from the light cylinder, that is for shock locations 
$\rm R_s >> R_{LC}$, where $\rm R_{LC}=8.1\times10^6\,cm$ 
is the radius of the light cylinder,
the upstream magnetic field  is given as: 
$\rm B_1 \sim [\sigma/(1+\sigma)]^{1/2}\, (\dot E /c\,f_p)^{1/2}\,R_s^{-1}$, where
$\rm f_p$ is the pulsar isotropic factor and $\rm \dot E$ is the
pulsar spin-down power \citep{Arons_Tavani93}.
J1227 has a powerful pulsar with 
$\rm \dot E \sim 9\times10^{34}\,erg\,s^{-1}$  and for $\rm R_s \lesssim 
R_{max} \sim 0.45\,a = 7\times10^{10}\,cm$ and for an isotropic wind,  
$\rm f_p$=1, we obtain  $\rm B_1 \gtrsim$ 1.4\,G for
$\sigma=3\times\,10^{-3}$  and $\rm B_1 \gtrsim 20-25$\,G for  
high-$\sigma$ wind conditions. 
Consequently,  $\rm B_2 \sim B_1 \gtrsim 20-25\,G$  for $\sigma>>1$,  
whilst it is  $\rm B_2 \sim 3 B_1 \gtrsim$ 4\,G for low-$\sigma$ regimes.
In both cases the magnetic field is large and higher than those in PWNe 
\citep{Kennel_Coroniti84}. 
These values are slightly larger than those derived by \cite{deMartino15}, 
who instead assumed the shock at the NS Roche lobe radius.
The Lorentz factor of accelerated electrons is related to the
maximum energy of synchrotron photons and post-shock magnetic field. The
{\it NuSTAR} spectrum does not show a break up to 70\,keV and thus we conservatively 
assume $\rm E_{x,max}$ at this energy.
Thus $\gamma \sim 2.5\times10^5\,\rm (E_{x,max}/B_2)^{1/2} \sim (5-10)\times10^5$ 
for high and low magnetizations, respectively  \citep{Rybicki_Lightman79}. 
The  electron population has a power law
energy spectrum with index $\rm p$ related to the X-ray power-law index 
$\Gamma$ as
$\rm p  \sim 2\Gamma -1 \sim$ 1.3  below the maximum 
energy, extending at least down to $\sim$0.1-0.3\,keV. 
Observations at lower energies
only encompass FUV and UV bands with the shortest wavelength coverage at 
2216\,$\AA$ \citep{Rivera_Sandoval18}. At these wavelengths an excess of 
flux over the companion star emission
has been detected primarly at orbital maximum  and ascribed to the possible 
contribution of the IBS in the UV \citep{Rivera_Sandoval18}. 
However the  extrapolation 
of the synchroton spectrum, using the derived power-law $\Gamma$ index,  
gives a flux 2-orders of magnitude lower than that observed at 2216\,\AA. 
Here we note that an additional bremsstrahlung 
component from an  ADAF-like wind inflow contributing to the unmodulated flux 
has been recently suggested at soft X-ray energies, 
extending to UV and optical wavelengths
\cite{Wadiasingh18}, that could be possibly related to the observed UV excess. 
Hence, we can only  set an upper 
limit of 0.1\,keV to the synchrotron energy of the minimum
Lorentz factor $\rm \gamma_{min}$, giving $\rm \gamma_{min}\lesssim 2-5\times 10^4$.
This range of high-$\gamma$ values implies a 
population of non-thermal emitting
electrons from the  shock in the X-ray domain.
The hard p-index appears to contradict
simple Fermi-type acceleration (diffusive shock acceleration) 
that predicts p$\sim$2.1-2.2 and could be in favour of 
magnetic dissipation (shock-driven magnetic reconnection) 
in the striped pulsar wind  that predicts p$\sim$1-2 in 
high-$\sigma$ regimes \citep[see][]{Sironi11a,Sironi11b,Sironi15}.

 The X-ray light curves of J1227 observed at different epochs all show a maximum
at inferior conjunction of the pulsar $\rm \phi_{orb}\sim 0.75$. This phasing is observed in all RBs and
it is opposite to what observed in BWs, that instead have a maximum at superior conjunction
of the pulsar. The different behaviour is explained with a shock that
wraps the pulsar in the RBs while the opposite occurs in BWs \citep{Romani16,Wadiasingh17}.
The light curves show different shapes over the years, with a 
broad asymmetric
maximum in 2013 and 2015 centred at $\rm \phi_{orb}\sim 0.75$ but peaking 
around $\rm \phi_{orb}\sim$0.7 and $\sim$0.9, respectively,
with similar amplitudes. Instead, the modulation increases by a factor $\sim$2.2
in 2014 and has a well defined  double-peaked maximum, 
with peak separation $\Delta\Phi_{orb}\sim 0.36$ and a dip reaching 
$\Delta I \sim 13\%$ with respect to the maximum. 
If the shocked pulsar wind retains a moderately relativistic bulk motion, 
the synchrotron emission is expected to be Doppler 
boosted at inferior conjunction and de-boosted
at superior conjunction of the NS, thus producing an X-ray orbital modulation 
\citep{Arons_Tavani93,Dubus10,Dubus15}. The bulk Lorentz factor 
$\rm \Gamma_L = (1-\beta^2)^{-1/2}$ and
bulk velocity $\beta=v/c$, drive the Doppler boosting as 
$\rm \delta_{boost} = 
[\Gamma_L (1- \beta\,cos\,\theta_v)]^{-1}$ where $\rm \theta_v$ is 
the viewing angle
of the observer \citep{Dubus10}. The synchroton flux is expected to be enhanced
by a factor $\rm \delta_{boost}^{2+(p-1)/2}$, along the orbit 
\citep{Dubus15} with a maximum to minimum flux ratio simplified as  
$\rm \sim [(1+\beta\,sin\,i)/(1-\beta\,sin\,i)]^{2+(p-1)/2}$ \citep{Dubus10}. 
The observed
modulation  amplitudes in 2013 and 2015, assuming $\rm (p-1)/2=\Gamma_x=1.2$ 
and  binary inclination
in the range $\rm 46^o-55^o$,  would then give $\beta\sim 0.2$ and 
$\rm \Gamma_L \sim 1.02$ while in  2014 $\beta\sim 0.3$ and 
$\rm \Gamma_L \sim 1.05$, 
indicating moderately relativistic bulk flow. A maximum flux enhancement of 
$\rm \sim 2\,\Gamma_{L,max}^{2+(p-1)/2}$ at the surface cap of the shock 
is predicted by the semi-analytical model 
of \cite{Wadiasingh17} implying $\rm \Gamma_{L,max} \sim$1.1 and $\sim$1.3 and 
$\beta_{max}\sim$ 0.4 and $\sim 0.7$ in the
observations of 2013/2015 and 2014, respectively. For moderately low binary inclinations
as in J1227,  broad maxima and weak amplitudes are predicted for low
$\rm \beta$ values and small shock radii, while 
distinct double-peaked maxima and large amplitudes are expected for high 
bulk velocities and larger shock distances from the pulsar \citep{Wadiasingh17}. 
The dip between the double-peaked maxima will have fractional intensities increasing
at higher $\beta$ and/or larger shock distances. Hence the broad maximum 
observed in 
2013 and 2015 indicates lower bulk velocities and smaller 
shock radii than those  in 2014.
Then, if the changes in the shape of the X-ray modulation are signatures 
of changes in the shock location, this may suggest that
in 2013 and 2015 the donor star wind momentum was larger 
(smaller shock distance from the pulsar). 
Since optical modulation  was stronger in 2013 and 2015 than  in 2014 
and anti-correlated with X-rays, it may be possible, 
as proposed in \cite{deMartino15}, that the shock had moved farther 
from the donor star at those two epochs, leaving more visible area of the heated 
face. The possibility that the companion Roche-lobe filling factor could
have changed, while appearing a viable explanation, would also imply a change in 
the optical flux at the orbital minimum, where the un-heated face of the
secondary contributes. The U-band flux is however similar in all observations 
within uncertainties at this phase.

\noindent In addition, the light curves from soft to hard X-rays show  
a minimum  skewed towards the rise to maximum. This is more evident in 2014 when
the modulation is stronger and where the multi-sinusoidal fits require
up to three harmonics. Although these are purely phenomenological descriptions, 
they may indicate that the structure of the IBS  is asymmetric
either due to Coriolis effects near the stagnation point \citep{Wadiasingh17} or
at the shock boundaries \citep{Wadiasingh18}.
Consequently also the irradiation of the secondary star is expected to be asymmetric
\citep[][]{Romani16,Kandel19}. The optical multi-color photometry of J1227
obtained in early 2015 indeed showed strong asymmetries in the
portion of the light curves rising to the maximum \citep{deMartino15}, 
supporting the possibility that an asymmetric shock may provide at least
some of the heating. 

\noindent Whether the shock is stable over years-timescale is still an unsolved 
matter since quasi-spherical radial infall on a pulsar is unstable on dynamical timescales
for distances outside the light cyclinder \citep[see][]{Burderi01}. 
Recently \citep{Wadiasingh18} have investigated two possible mechanisms 
that could provide stability of a configuration where the shock is curved around the pulsar.
One where the secondary star has a large dipole magnetic field of several kilogauss
and low mass loss rates ($\rm \lesssim 10^{15}\,g\,s^{-1}$) and the other where 
instead the secondary mass loss rate is large ($\rm \sim 10^{15}-10^{16}\,g\,s^{-1}$) 
with the flow in an ADAF-like configuration. However, both mechanism have their 
shortcomings \citep[see details in][]{Wadiasingh18}.\\

The lack of a detectable orbital modulation above 25\,keV is unexpected result,  
not reported before for any of the other four RBs observed with {\em NuSTAR}, namely
PSR\,J1023+0038 \citep{tendulkar14}, PSR\,J2129-0429 \citep{AlNoori18}, PSR\,J1723-2837
\citep{kong17} and PSR\,J2339-0533 \citep{Kandel19}, although this could well
depend on the choice of the energy bands used to study the X-ray modulations. Indeed, 
a more detailed study of the energy resolved  X-ray orbital modulation in 
PSR\,J1023+0038 as observed with {\em NuSTAR} during its previous rotation-powered
state reveals that above $\sim$25\,keV the modulation has a fractional 
amplitude $\sim 15\pm3\%$, lower than that below 25\,keV,  $\sim 25\pm2\%$ (1$\sigma$
uncertainty) (Coti Zelati et al., in prep.). This smaller amplitude is 
consistent 
with the 3$\sigma$ upper limit of $\sim 10\%$ found in J1227. Small changes
in the spectral shape between the minimum and maximum of the orbital modulation,
although here undetected, could produce energy dependent modulation 
amplitudes.  
For PSR\,J2339-0533
\cite{Kandel19} find subtle changes in the phase-resolved X-ray
spectra and a double-peaked
light curve only below 15\,keV, while between 15-40\,keV it is single peaked. 
We here note that also for J1227, if studied over this wider energy range 
(e.g. 15-40\,keV) the orbital modulation has about similar amplitude as in the softer 
{\it NuSTAR}  bands. 
Indications of spectral changes along the double-peaked orbital modulation
were also found in the {\it XMM-Newton} and {\it NuSTAR}  data of 
PSR\,J2129-0429 \citep{AlNoori18}. 
These would be the signature of different populations of cooling electrons 
whose spectrum is sensitive to the shape of the shock geometry and on the 
downstream magnetic field along the shock \citep{Wadiasingh17,Kandel19}. 
If the orbital modulation has a lower amplitude at high energies  
this may indicate that the electron population contributing to the 
shock emission at these energies has a different p-index
or that a spectral break is 
hidden by the underlying magnetospheric pulsar emission.
The latter has been successfully described by a magnetospheric synchrocurvature
radiation model for the non-thermal $\gamma$ and X-ray spectra of pulsars
\citep{Torres18,Torres19} and would be dominant at minimum of the
orbital modulation. The lack of detectable phase-resolved 
spectral variations above 3\,keV (see Table.\,\ref{spec_fit}). 
cannot help in discriminating the 
spectral shapes of the magnetospheric and IBS emissions. In fact
even including two power-law components in the spectral fits and fixing
the parameters of one to those found for the minimum we derive  
similar power-law indexes within errors and larger normalizations
at the maximum, at the rise/decay to/from maximum. 

\noindent For what concerns the softer X-ray band,
i.e. 0.3-3\,keV, as observed by {\em XMM-Newton} 
the modulation amplitude is lower than that at 3-6\,keV by a factor 
of $\sim$ 1.4 in 2013 and 2014 (see Table\,\ref{sinu_fits}). 
This would produce an overall hardening at orbital maximum or, 
viceversa a softer emission at orbital minimum and, indeed, 
 a marginal change of the
power-law index  $\Gamma$ from $\sim$1.1  to $\sim$1.3 between maximum
and minimum was found \citep{deMartino15}.
Also in this case, an energy dependence of the X-ray light curve could 
be a signature of a 
spatial dependence along the shock of the electron p-index and 
thus less efficient (de-boosting) acceleration \citep{Wadiasingh17,Wadiasingh18}. 
However the lack of spectral 
change and thus of energy dependent modulation above
3\,keV makes this hypothesis less viable.
Alternatively, the softening at orbital minimum, i.e. at superior conjunction
of the pulsar, below 3\,keV could be explained by the additional 
contribution of the thermal 
emission of the NS where it is expected to be best visible.
In \cite{deMartino15} an upper limit of 
$\sim3\%$  of the NS contribution to the average 0.3-10\,keV flux was found,
while at orbital minimum it is $\lesssim 40\%$.
Although this does not allow  a firm conclusion, it appears a viable 
interpretation. A future  much longer
exposure in the soft X-ray band may allow to put more stringent constraints
on the NS thermal emission. 

A strong irradiation (from $\sim$5500\,K to 6500\,K) of the companion star 
in J1227 was found from optical spectroscopy
and photometry, persistent in both rotation-powered and disc-states 
\citep{deMartino14,deMartino15}. 
Whether the late-type secondary is heated by
the pulsar wind, by the X-ray emission from the IBS o both, 
could not be assessed due to the limited energy coverage of    
previous {\em XMM-Newton} observations. 
With the {\em NuSTAR} observation presented here, we derived  a   
X-ray luminosity  $\rm L_x \sim 10^{33}\,erg\,s^{-1}$, implying an efficiency
$\rm \eta_x = L_x/\dot E \sim 1\%$. 
The expected heating power 
impinging onto the companion star in the case of
an isotropic pulsar wind is 
 $\rm L_{heat,SD} = f_{\Omega}\,\dot E$ where $\rm f_{\Omega}$ is
the geometric factor  $\rm \sim 0.5\,(1- cos\,\Omega)$ with 
$\rm \Omega = atan(R_2/a)$, neglecting albedo. 
From optical pho\-to\-me\-try \citep{deMartino15}, the secondary is
found to fill its Roche lobe and thus 
$\rm R_2/a=R_{L2}/a=0.462\,(q/1+q)^{1/3}$,  giving 
$\rm f_{\Omega} \sim 0.01$ for q=0.194. We then derive 
$\rm L_{heat,SD} \sim 9\times 10^{32}\,erg\,s^{-1}$. 
Similarly, if we consider the X-ray IBS emission located at
$\sim 0.45\,a$ from the MSP, we obtain 
$\rm L_{heat,X} \sim 4 \times10^{31}\,erg\,s^{-1}$, which is only
$\sim 4.4\%$ the expected heating from the pulsar spin-down power, 
and is only $\sim 3\%$  
the luminosity of the irradiated face of the companion, 
$\rm L_{irr}=4\,\pi\,R_2^2\, \sigma_{SB}\,T_{irr}^4 \sim 1-2\times 
10^{33}\,erg\,s^{-1}$ for $\rm T_{irr}\sim$ 6500\,K
\citep{deMartino15}\footnote{The effect of irradiation is to
suppress the tem\-pe\-ra\-tu\-re gra\-dient in the outer stellar envelope, 
blo\-cking the out\-wards transport of energy through the face. A detailed rigorous 
treatment of irradiation can be found in \cite{Ritteretal00}}. 
Then the total irradiating luminosity, $\rm L_{heat,SD} + L_{heat,X}
\sim 10^{33}\,erg\,s^{-1}$,  is just about what necessary to
heat the companion star. Different is the case where the shock power-law
extends into the MeV range ($\sim$10\,MeV), in which case the IBS 
alone would be energetically able to irradiate the secondary star. 
Alternatively the shock is located much closer (at) the companion 
but this would contrast with the geometry of the IBS 
wrapping the pulsar for the correct phasing of the orbital modulation.

\noindent We also note that  additional heating could originate 
from shock particles that thread the companion magnetic field lines
and are ducted to its surface \citep{Sanchez17}. Such possibility
would require kilogauss magnetic field strengths for the secondary,
hence a very active magnetic late-type star (large star-spots and/or 
flares) \citep{Wadiasingh18,Kandel19}. So far large star-spots
are not identified in the optical light curves of RBs, although a few
systems have been found to display flares \citep{Romani15,Deneva16,Cho18}.
This additional heating has been invoked to explain asymmetries in the
light curves of many RBs \citep{Romani15,Sanchez17}.  J1227 could also
be affected by this mechanism.

\section{Conclusions}

We have presented the first {\em NuSTAR} hard X-ray data of the
tMSP J1227 during its rotational-power state complemented with coordinated a
2.5\,yr-long radio monitoring at {\em Parkes} telescope and with
archival {\em XMM-Newton} and {\em Swift} data.
Here we summarise the main results:

\begin{itemize}
\item
 The {\em Parkes} monitoring gives us a refined orbital
radio ephemeris around the {\em NuSTAR} observation, which we used
to study the orbital dependent  X-ray emission.
The radio observations, mainly conducted at 1.4\,GHz confirm the presence
of long eclipses lasting $\sim40\%$ of the orbit. The distance to 
J1227 derived from the DM is consistent with that obtained from {\em Gaia} 
DR2 parallax, D=$1.37_{-0.15}^{+0.69}$\,kpc.

\item
The X-ray spectrum is non-thermal extending  up to $\sim$70\,keV 
withouth a spectral break and is consistent with a power-law, with
photon index $\Gamma=1.17\pm$0.08. The X-ray emission is ascribed
to an  intrabinary shock  formed between the pulsar and companion winds.
We derive a luminosity in the 3-79\,keV range of 
$\rm 7.6_{-0.8}^{+3.8}\times
10^{32}\,erg\,s^{-1}$, using the {\em Gaia} distance.

\item
The X-ray emission in the 3-79\,keV range displays significant 
modulation at the 6.91\,h orbital period with an amplitude of 28$\pm 3\%$.
Energy resolved orbital light curves show that the modulation
is significantly detected below $\sim$25\,keV with no energy dependence.
Comparison with previous soft X-ray observations in the 0.3-12\,keV
with {\em XMM-Newton} reveals a mild softening only below 3\,keV at
superior conjunction of the NS. This suggests the possible 
contribution of the NS atmosphere to be confirmed with deeper
soft X-ray observations than acquired so far.

\item
We derive a 3$\sigma$ upper limit to the modulation fraction 
above 25\,keV of $10\%$, which suggests that the 
electron population contributing to the shock emission at 
these energies is subject to  different acceleration. 

\item
The folded orbital light curves in common energy ranges as observed 
over three epochs
by {\em NuSTAR} and {\em XMM-Newton} reveal changes in the amplitude 
over the years, indicating that the shock is not stationary
and varies with time since J1227 transitioned into a rotation-powered 
state.
We confirm previous finding of an anti-correlated variability 
between X-ray and optical ranges in the orbital modulation amplitudes.

\item
We estimate for the shock  luminosity in the  0.3-79\,keV range  
$\rm 8.3_{-0.9}^{+4.2}\times10^{32}\,erg\,s^{-1}$, that is
not enough alone to power the irradiation of the companion star unless
the emission extends to the MeV range or the shock does not 
wrap the pulsar. 

\item 
A search for X-ray pulses in the 3-79\,keV range provides us
with an upper limit to the rms pulse amplitude of
$8.8\%$, compatible with the upper limit derived in the 0.3-10\,keV  band 
from previous {\em XMM-Newton} data. 

\end{itemize}
 
\section*{Acknowledgments}
This work is based on observations obtained with the NASA science 
mission {\em NuSTAR}, with {\em XMM-Newton}, an ESA science mission 
with instruments and 
contributions directly funded by ESA Member States, with 
{\em Swift}, a NASA science mission with Italian participation,
with {\em Parkes} telescope and  
with \textit{Gaia}, an ESA mission. \textit{Gaia} data are 
processed by the Data Processing and Analysis Consortium 
(DPAC).
This work made use of data supplied by the UK {\em Swift} Science Data 
Centre at the University of Leicester. 
DdM, AP and TMB acknowledge financial support from the Italian Space Agency 
(ASI) and National Institute for Astrophysics (INAF) under agreements 
ASI-INAF I/037/12/0 and ASI-INAF n.2017-14-H.0 and from INAF "Sostegno alla
ricerca scientifica main streams dell'INAF", Presidential Decree 43/2018. 
DdM, AP, MB and AP 
also acknowledge support from INAF "SKA/CTA projects", 
Presidential Decree N. 70/2016.
DFT acknowledges support from the grants PGC2018-095512-B-I00, 
SGR2017-1383, and AYA2017-92402-EXP. 
FCZ is supported by a Juan de la Cierva fellowship and grants 
SGR2017-1383 and PGC2018-095512-B-I00. We also thank the ''PHAROS'' COST Action 
(CA16214) for partial support. 
We also acknowledge useful discussion with Dr. A.\,Sanna on the {\em NuSTAR} 
data timing analysis.

\bibliographystyle{mnras}
\bibliography{xss_nustar_accepted.bbl}

\label{lastpage}

\vfill\eject

\end{document}